\PassOptionsToPackage{table}{xcolor}
\documentclass[10pt, final, twocolumn, twoside, romanappendices]{IEEEtran}
\usepackage{amsmath,amssymb}
\usepackage[level=0]{LatexInclusion/wgroup_message}  

\usepackage{bm}

\DeclareMathAlphabet{\mathsfbr}{OT1}{cmss}{m}{n}
\SetMathAlphabet{\mathsfbr}{bold}{OT1}{cmss}{bx}{n}
\DeclareRobustCommand{\msf}[1]{%
  \ifcat\noexpand#1\relax\msfgreek{#1}\else\mathsfbr{#1}\fi
}

\makeatletter
\newcommand{\msfgreek}[1]{\csname s\expandafter\@gobble\string#1\endcsname}
\makeatother

\DeclareFontEncoding{LGR}{}{} 
\DeclareSymbolFont{sfgreek}{LGR}{cmss}{m}{n}
\SetSymbolFont{sfgreek}{bold}{LGR}{cmss}{bx}{n}
\DeclareMathSymbol{\salpha}{\mathord}{sfgreek}{`a}
\DeclareMathSymbol{\sbeta}{\mathord}{sfgreek}{`b}
\DeclareMathSymbol{\sgamma}{\mathord}{sfgreek}{`g}
\DeclareMathSymbol{\sdelta}{\mathord}{sfgreek}{`d}
\DeclareMathSymbol{\sepsilon}{\mathord}{sfgreek}{`e}
\DeclareMathSymbol{\szeta}{\mathord}{sfgreek}{`z}
\DeclareMathSymbol{\seta}{\mathord}{sfgreek}{`h}
\DeclareMathSymbol{\stheta}{\mathord}{sfgreek}{`j}
\DeclareMathSymbol{\siota}{\mathord}{sfgreek}{`i}
\DeclareMathSymbol{\skappa}{\mathord}{sfgreek}{`k}
\DeclareMathSymbol{\slambda}{\mathord}{sfgreek}{`l}
\DeclareMathSymbol{\smu}{\mathord}{sfgreek}{`m}
\DeclareMathSymbol{\snu}{\mathord}{sfgreek}{`n}
\DeclareMathSymbol{\sxi}{\mathord}{sfgreek}{`x}
\DeclareMathSymbol{\somicron}{\mathord}{sfgreek}{`o}
\DeclareMathSymbol{\spi}{\mathord}{sfgreek}{`p}
\DeclareMathSymbol{\srho}{\mathord}{sfgreek}{`r}
\DeclareMathSymbol{\ssigma}{\mathord}{sfgreek}{`s}
\DeclareMathSymbol{\stau}{\mathord}{sfgreek}{`t}
\DeclareMathSymbol{\supsilon}{\mathord}{sfgreek}{`u}
\DeclareMathSymbol{\sphi}{\mathord}{sfgreek}{`f}
\DeclareMathSymbol{\schi}{\mathord}{sfgreek}{`q}
\DeclareMathSymbol{\spsi}{\mathord}{sfgreek}{`y}
\DeclareMathSymbol{\somega}{\mathord}{sfgreek}{`w}

\DeclareMathSymbol{\svarsigma}{\mathord}{sfgreek}{`c}

\DeclareMathSymbol{\sGamma}{\mathalpha}{sfgreek}{`G}
\DeclareMathSymbol{\sDelta}{\mathalpha}{sfgreek}{`D}
\DeclareMathSymbol{\sTheta}{\mathalpha}{sfgreek}{`J}
\DeclareMathSymbol{\sLambda}{\mathalpha}{sfgreek}{`L}
\DeclareMathSymbol{\sXi}{\mathalpha}{sfgreek}{`X}
\DeclareMathSymbol{\sPi}{\mathalpha}{sfgreek}{`P}
\DeclareMathSymbol{\sSigma}{\mathalpha}{sfgreek}{`S}
\DeclareMathSymbol{\sUpsilon}{\mathalpha}{sfgreek}{`U}
\DeclareMathSymbol{\sPhi}{\mathalpha}{sfgreek}{`F}
\DeclareMathSymbol{\sPsi}{\mathalpha}{sfgreek}{`Y}
\DeclareMathSymbol{\sOmega}{\mathalpha}{sfgreek}{`W}

\DeclareRobustCommand{\mcal}[1]{%
  \ifcat\noexpand#1\relax\mathnormal{#1}\else\cal{#1}\fi
}
\DeclareRobustCommand{\BM}[1]{%
  \ifcat\noexpand#1\relax\bm{\boldUppercaseItalicGreek{#1}}\else\bm{#1}\fi
}
\makeatletter
\newcommand{\boldUppercaseItalicGreek}[1]{\csname var\expandafter\@gobble\string#1\endcsname}
\makeatother

\newcommand{\V}[1]{\bm{#1}}
\newcommand{\M}[1]{\BM{#1}}

\hypersetup{colorlinks,citecolor= red,filecolor= blue,linkcolor= blue,urlcolor=blue}
\usepackage{acronym}  
\usepackage{color}  
\usepackage{graphicx}%
\usepackage[usenames,dvipsnames]{xcolor}
\usepackage{float}
\usepackage{multirow}
\usepackage{comment}
\usepackage{soul}
\usepackage{subcaption}
\usepackage[many]{tcolorbox}
\usepackage{xcolor}
\usepackage[export]{adjustbox}

\usepackage{textcomp}

\newcommand{\bd}{\begin{description}}
\newcommand{\ed}{\end{description}}
\newcommand{\be}{\begin{enumerate}}
\newcommand{\ee}{\end{enumerate}}
\newcommand{\bi}{\begin{itemize}}
\newcommand{\ei}{\end{itemize}}
\newcommand{\bl}{\begin{list}}
\newcommand{\el}{\end{list}}
\newcommand{\bt}{\begin{tabbing}}
\newcommand{\et}{\end{tabbing}}

\definecolor{BLUE}{rgb}{0,0,1}

\usepackage{acronym}
\acrodef{ook}[OOK]{on-off keying}
\acrodef{rc}[RC]{resistor-capacitor}
\acrodef{ac}[AC]{alternating current}
\acrodef{dc}[DC]{direct current}
\acrodef{ber}[BER]{bit error rate}
\acrodef{i2v}[I2V]{infrastructure-to-vehicle}
\acrodef{v2v}[V2V]{vehicle-to-vehicle}
\acrodef{led}[LED]{light emitting diode}
\newacro{vlc}[VLC]{visible light communication}
\acrodef{uwb}[UWB]{ultra-wideband}
\acrodef{fy}[FY]{fiscal year}

\acrodef{iot}[IoT] {Internet of Things}
\newacro{aws}[AWS]{Amazon Web Services}

\acrodef{ble}[BLE]{Bluetooth Low Energy}

\newacro{dfe}[DFE]{decision feedback equalizer}
\newacro{lms}[LMS]{least means square}
\newacro{ann}[ANN]{artificial neural network}
\newacro{alu}[ALU]{arithmetic logic unit}
\newacro{api}[API]{application programming interface}
\newacro{asic}[ASIC]{application-specific integrated circuit}

\newacro{sdn}[SDN]{software defined network}
\newacro{hca}[HCA]{heterogeneous computing architecture}

\acrodef{gps}[GPS]{Global Positioning System}
\acrodef{imu}[IMU]{inertial measurement unit}
\acrodef{uav}[UAV]{unmanned aerial vehicles}

\acrodef{csi}[CSI]{channel state information}
\acrodef{snr}[SNR]{signal-noise-ratio}

\acrodef{los}[LOS]{line-of-sight}
\acrodef{nlos}[NLOS]{non-line-of-sight}
\newacro{toa}[TOA]{time-of-arrival}
\newacro{tdoa}[TDOA]{time-difference-of-arrival}
\newacro{aoa}[AOA]{angle-of-arrival}
\newacro{rss}[RSS]{received signal strength}

\newacro{crb}[CRB]{Cram\'{e}r-Rao Bound}
\newacro{speb}[SPEB]{squared position error bound}
\newacro{fim}[FIM]{Fisher information matrix}
\newacro{pocs}[POCS]{projection onto convex sets}

\newacro{eed}[EED]{Engineering Every Day}
\newacro{stem}[STEM]{science, technology, engineering, and mathematics}

\usepackage[yyyymmdd,hhmmss]{datetime} 
\newcommand{\versionnumber}{11 R1}
\newcommand{\version}{\textcolor{BLUE}{Version: V\versionnumber -- \textsc{\monthyeardate\today}}}

\newdateformat{monthyeardate}{\monthname[\THEMONTH] \THEDAY, \THEYEAR} 

\newcommand{\paperTitle}{
	Measurement-based VLC channel characterization for I2V communications in a real urban scenario
}

\newcommand{\paperTitleMarkboth}{Measurements-based VLC channel characterization for I2V communications}

\newcommand{\Abstract}{Visible light communication (VLC) is nowadays envisaged as a promising technology to enable new classes of services in intelligent transportation systems ranging, e.g., from assisted driving to autonomous vehicles. The assessment of the performance of VLC for automotive applications requires as a basic step a model of the transmission pattern and propagation of the VLC signal when real traffic-lights and road scenarios are involved. In this paper an experimental measurement campaign has been carried out by using a regular traffic-light as source \textcolor{black}{(red light)} and a photoreceiver positioned, \textcolor{black}{statically}, at different distances and heights along the road. A linear regression technique is used to come up with different propagation models. \textcolor{black}{The proposed models have been compared, in terms of accuracy and complexity, to the conventional Lambertian model to describe the VLC channel in a real urban scenario. The proposed models provides a significant higher accuracy with comparable complexity.}}

\newcommand{\keywords}{	
	Visible light communications, experimental measurements, channel modeling, vehicular communications.
}

\begin{document}
	\twocolumn
	\title{\paperTitle}
	\author{
		\vspace{0.2cm}
		S. Caputo,
        L. Mucchi,
		F. S. Cataliotti, M. Seminara, T. Nawaz and
        J. Catani
		 \\
		\vspace{0.4cm}
 		\vspace{-0.2cm}

\thanks{\textcopyright 2019 IEEE. Personal use of this material is permitted. Permission from IEEE must be obtained for all other uses, in any current or future media, including reprinting/republishing this material for advertising or promotional purposes, creating new collective works, for resale or redistribution to servers or lists, or reuse of any copyrighted component of this work in other works.

\vspace{0.25cm}

Stefano Caputo and Lorenzo Mucchi are with the Dept. of Information Engineering (DINFO), University of Florence, Italy (e-mail: \texttt{stefano.caputo@unifi.it; lorenzo.mucchi@unifi.it}). 

Francesco Cataliotti is with the European Laboratory for Non linear Spectroscopy (LENS) and Dept. of Physics and Astronomy, University of Florence, Italy (e-mail: \texttt{fsc@lens.unifi.it}).

Marco Seminara is with the European Laboratory for Non linear Spectroscopy (LENS) and National Institute of Optics-CNR (CNR-INO), Sesto F.no, Italy (e-mail: \texttt{seminara@lens.unifi.it}).

Tassadaq Nawaz is with the National Institute of Optics-CNR (CNR-INO), Sesto Fiorentino, Italy
(e-mail: \texttt{tassadaq.nawaz@ino.it}).

Jacopo Catani is with the National Institute of Optics-CNR (CNR-INO) and LENS, Sesto Fiorentino, Italy (e-mail: \texttt{jacopo.catani@ino.it}). 
		}
	}
	\maketitle
	\markboth{Submitted to IEEE Trans. on Veh. Tech. \version}
	{Caputo, et al.: \paperTitleMarkboth}
	\setcounter{page}{1}
	\begin{abstract}
		\Abstract
	\end{abstract}
	
	\acresetall		


\section{Introduction}\label{sec:intro}
The ongoing substitution of conventional light sources with light emitting diodes (LEDs) has recently fuelled scientific and industrial activity in \ac{vlc} technology \cite{KomineNakagawa,Alfattani,ChiHaas}. Such an interest stems from the possibility offered by LEDs for fast modulation of light. 
In the context of automotive and vehicular networks \cite{Martinez,Karagiannis}, where low latency and reliability are of crucial importance, \ac{vlc} can offer significant advantages \cite{WuWangYoun,Chi}. In addition, \ac{vlc} is an inherently energy-efficient interconnection since the energy is, in any case, already needed to illuminate the road or for road signaling (e.g., traffic lights) \cite{Akanegawa}. \ac{vlc} could therefore be significant for \ac{v2v} or \ac{i2v} communications, both issues of decisive importance for road safety, in particular in the context of assisted- or unmanned-driving \cite{Correa,BobanKousaridas}. 
A reliable assessment of suitability of \ac{vlc} technology for V2V or I2V communications \cite{Hu,BobanManolakis} requires an accurate characterization of the \ac{vlc} channel. 
In literature, the approaches to this problem has been theoretical or empirical. 
The theoretical approach aims at a mathematical reduction of the problem with approximations, such as Lambertian emission and reflection pattern \cite{Akanegawa}. Ray tracing is another method of investigation, allowing for a software-based physical simulation \cite{LeeKwon}. This kind of approaches is usually characterized by a simplified scenario, aimed at reducing the computational complexity. On the other hand, the empirical approach typically involves a measurement campaign followed by the extraction of a mathematical model of the channel as in \cite{CuiChen}.
However, whilst the optical wireless channel has been characterized in literature for infrared (IR) communication \cite{Barry}, and recently a comparison between IR and \ac{vlc} appeared \cite{LeePark}, in order to enable the use of \ac{vlc} technology for safety-critical / smart driving applications \cite{Yaqoob,Ayub}, it is mandatory to test and analyze the \ac{vlc}-based \ac{v2v} and \ac{i2v} communication channel in a real urban environment, with regulatory \acs{led} sources and infrastructures. \textcolor{black}{As for today, however, despite recent works demonstrating fast and efficient I2V communications up to $50$\,m using regulatory traffic lights \cite{Nawaz_2019_IEEE}\cite{8864129}, a comprehensive characterization of the \ac{vlc} channel in a realistic outdoor scenario is still lacking, due to intrinsic difficulties mainly represented by the influence of non-ideal external factors, such as ambient light and irregularities of the emission pattern of the \acs{led}-based traffic lights or headlights \cite{Moreira,Islim}}. 
In this paper we present an outdoor measurement campaign, aimed at the extensive characterization of the \ac{vlc} channel in a real urban scenario, carried out in the city of Prato (Italy) in collaboration with ILES srl, a private company which develops urban signaling systems. The measurements have been taken in a real urban road, with regulatory traffic light emitting a \ac{vlc} signal to the receiver located along the same road. 
The measurement grid has been replicated setting the receiver at three different heights corresponding, respectively, to car headlights, dashboard and internal mirror. Measurements have been performed also in the presence of stray light sources such as sunlight or other cars' headlights. 
The data has been used to extract a mathematical model of the \ac{vlc} signal propagation. In particular, the optical source plus propagation transfer function is modeled. We will show that the Lambertian model is far from being accurate when real traffic-light lamps are considered for \ac{vlc} transmission. A linear regression technique is used to come up with a more accurate model of the transmission-propagation pattern.   
The results of the measurement campaign highligts attainable distances of several tens of meters, mainly limited by the directional emission pattern of the semaphore lamps, optimized for maximum visibility at $\simeq 15\,\rm m$ for typical dashboard heights.
The key contributions of this paper can be summarized as follows: 
\begin{itemize}
	\item a specific hardware for feeding and modulating the current flowing into a red LEDs array of a regular traffic-light has been designed and implemented, along with a physically AC-coupled RX stage for DC, high-intensity stray components rejection;
    \item a measurement campaign of a \ac{vlc} link between a traffic-light and a receiving unit has been carried out in a real urban road, with regulatory traffic-light and environmental conditions; three different heights have been taken into account, corresponding to headlights, dashboard and internal mirror heights in a standard car; 
    \item a linear regression technique has been used to derive a mathematical model of the \ac{vlc} transmission pattern plus propagation;  
    \item The \ac{ber} of the \ac{i2v} \ac{vlc} system has been estimated from the measured \ac{snr} at the receiver. 
\end{itemize} 

The rest of the paper is organized as follows. Sec.\,\ref{sec:systemModel} introduces the complete system model, while Sec.\,\ref{sec:Measurement Campaign} shows the measurements campaign set up, including the hardware equipment.  In the following Sec.\,\ref{propagation model} different propagation models are discussed, derived from the experimental measurements. Sec.\,\ref{sec:Results} comments the results and Sec.\,\ref{sec:conclusion} concludes the paper. 
        


\section{System Model}\label{sec:systemModel}

\begin{figure}[tb]
			\centering
			\includegraphics[width=0.9\columnwidth]{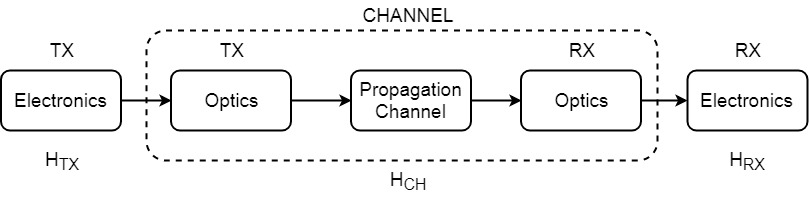}
			\caption{The VLC system model. The red dotted line indicates the blocks we are aiming to model.}
			\label{fig:blockdiagramTLC}
		\end{figure}

\begin{figure}[h]
	\centering
	\includegraphics[width=\columnwidth]{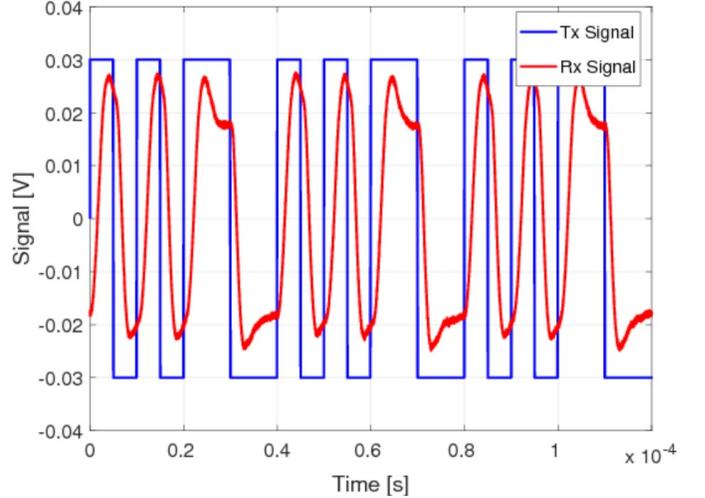}
    \caption{The blue curve represents the transmitted signal (TX), modulated with \ac{ook} Manchester NRZ scheme, as indicated in standard IEEE 802.15.7 for outdoor \ac{vlc}s. The red curve represents the signal received (RX) at the reference position indicated in Fig.\,\ref{fig:expsetup}, averaged over 4 acquisition periods. \color{black}The comparison highlights a small delay, $\sim$ 2 $\mu$s, related to the finite bandwidth of the TX-RX chain.\color{black}}
	\label{fig:signaltxrx}
\end{figure}

The VLC channel can be decomposed in three elements: two corresponding to the electronic circuits, i.e., the transmitter and the receiver, and an optical channel. The optical channel can again be decomposed in three elements: two corresponding to the optics installed on the transmitter and on the receiver, respectively, and one corresponding to propagation in free space. Our model for the VLC channel is shown in Fig.\,\ref{fig:blockdiagramTLC}.
Every single element of the scheme can be represented mathematically by its transfer function $H(f)$, corresponding to the Fourier Transform of the impulse response $h(t)$.  
In our implementation, the signal transmitted by the LED lamp is represented by a time-dependent rectangular function
\begin{equation}\label{eq.s(t)}
    s(t) = \sum_{j=0}^{\infty} s_j(t-jN_bT) 
\end{equation}
where 
\begin{equation}\label{eq.waveform}
    s_j(t) = A \sum_{i=0}^{7} s_k \cdot \text{rect}\left(\frac{t-T/2+kT}{T} \right)
\end{equation}
and $A$ is the amplitude, $T$ is the duration of the single rectangular pulse, $s_k$ is the transmitted symbol and $N_b$ is the number of transmitted symbol sequences. The symbol sequence is $\V{s}=\{1, -1, 1, -1, 1, 1, -1, -1\}$, chosen in order to encompass all of the possible logic transitions appearing in Manchester encoding, where no more than two consecutive symbols can be of the same sign \cite{Dahri} (see Fig.\,\ref{fig:signaltxrx}). 

The received signal, in each point on the grid, is the convolution of the transmitted signal with four impulse responses 
\color{black}
\begin{equation}\label{eq:reception}
    s_{\text{RX}}(t) = s_{\text{TX}}(t) * h_{\text{TX}}^{\text{el}}(t) * h_{\text{TX}}^{\text{op}} * h_{\text{P}}(t) * h_{\text{RX}}^{\text{op}} * h_{\text{RX}}^{\text{el}}(t) + n(t)
\end{equation} 
\color{black}
where 
\begin{itemize}
    \item \textcolor{black}{$n(t)$ is the (white Gaussian) thermal noise at the receiver},  
    \item $h_{\text{TX}}^{\text{el}}(t)$ takes into account the electronic at the transmitter, 
    \item $h_{\text{TX}}^{\text{op}}$ the effects of the optics at the transmitter, 
    \item $h_{\text{P}}(t)$ the light propagation, 
    \item $h_{\text{RX}}^{\text{op}}$ the optics at the receiver, and 
    \item $h_{\text{RX}}^{\text{el}}(t)$ the electronic conversion at the receiver. 
\end{itemize}
In the frequency domain, Eq.~\eqref{eq:reception} becomes 
\color{black}
\begin{equation}\label{eq:rec-freq}
    S_{\text{RX}}(f) = S_{\text{TX}}(f) \underbrace{H_{\text{TX}}^{\text{el}}(f) H_{\text{TX}}^{\text{op}} H_{\text{P}}(f) H_{\text{RX}}^{\text{op}} H_{\text{RX}}^{\text{el}}(f)}_{H(f)} + N(f)
\end{equation}
\color{black}

It is important to note that $h_{\text{P}}(t)$ does not actually depend on time as well, since during the measurements both TX and RX were static, and no variations in the environment occurred. 
In our study, we aim at the characterization of ultimate performance of our system in a real scenario. For this motivation, for each point of the grid, we chose to align the optical axis of the receiver towards the lamp, adjusting its horizontal and vertical alignment angles. The benefits of such setting are mainly two. First, with our set of focal length and detector size, the image of the lamp falls entirely into the active area of the detector for all points on the grid, so that the reconstructed intensity pattern coincides with the lamp pattern. Secondly, this configuration avoids any interference effect given by reflections of the lamp's signal from tarmac, as no reflected beam falls into the field of view (FOV) of our receiver. A demonstration of this is given in Appendix\,\ref{app:fov}.

For the above consderations, under our configuration we can  assume $h_{\text{RX}}^{\text{op}}$ to be nearly constant into the whole measurement grid.
The impact of different collecting optics and angles on the overall system performances, as well as possible effects of reflections on the optical channel quality will be addressed in future work.

In this paper we focus on the modelling of $ h_{\text{TX}}^{\text{op}}* h_{\text{P}}(t)$. It is common in literature to model this as Lambertian, but we will demonstrate that the Lambertian model does not provide for an accurate description when the source is a real LED-based traffic-light, equipped with regulatory shaping optical lens, in an urban scenario. We then propose a linear regression method to come up with a mathematical model that better fits the propagation of VLC signals from traffic-light to vehicles. 


\section{Measurement Campaign in Urban Scenario}\label{sec:Measurement Campaign}
\begin{figure}[tb]
\centering
\includegraphics[width=0.9\columnwidth]{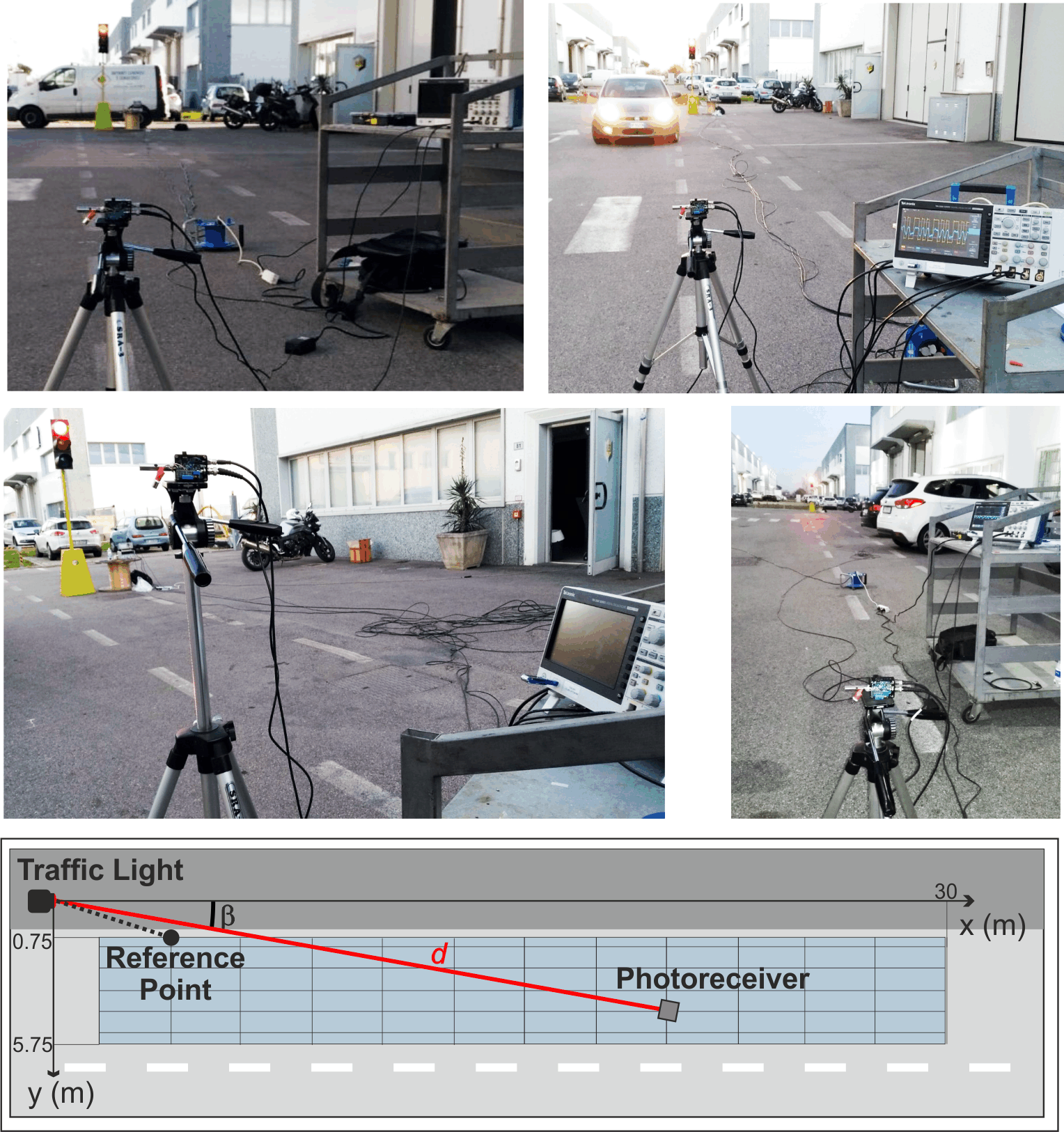}
\caption{Experimental setup for the measurement campaign in a real urban scenario. Several different conditions has been tackled. The AC coupling of the photodetector washes the contribution of ambient and artificial stray lights out during all the day and the evening, even in the case of direct stray sunlight and/or headlight illumination. In the lower panel, a sketch of the measurement grid is reported. Measurements have been repeated in the grid-area for three different receiver heights (see main text). The reference (0 dB) point is located at (4, 0.75) m, for a height of 1.35 m.}
\label{fig:expsetup}
\end{figure}

The on-field measurements campaign have been carried out in the city of Prato (Italy) in collaboration with ILES srl, a company producing and installing road signaling infrastructures. The company has installed one traffic light in a real urban scenario consisting of a two-lane road with buildings on both sides (see Fig.~\ref{fig:expsetup}). The traffic light has been positioned on the right side of the rightmost lane, with 0.75 m indentation, at the height of 2.83 m, in accordance to Italian regulations UNI11248 \cite{UNI11248} and UNI13201-2 \cite{UNI13201-2}. The measurements have been carried out during one entire day (sunny conditions). 
The red lamp of the traffic light has been modulated with the information signal (see Fig.~\ref{fig:expsetup}). In particular, the red LED lamp has been controlled by our driver circuit (see Sec.~\ref{sec:TX}) which, besides controlling and supplying the DC nominal operating current, allows for the insertion of the bit sequence via an external function generator (see Sec.~\ref{sec:TX}). The LED driver is the only component that has been replaced in the commercial traffic light . 

The photodetector has been positioned in a grid on points in front of the emitting traffic light (Fig.~\ref{fig:expsetup}). The grid has been centered at the position of the traffic light, with the $x$-axis along the road length, the $y$-axis along the road width and a vertical $z$-axis. The grid points along the x-axis ranges from 3 to 30\,m, with more density in the first 10\,m: [3\,m, 4\,m, 5\,m, 7\,m, 10\,m, 15\,m, 20\,m, 25\,m, 30\,m]. Along the y-axis we included an offset of 0.75\,m since the traffic light has this indentation on the sidewalk. The step between measurement points is fixed to 1\,m: [0.75\,m, 1.75\,m, 2.75\,m, 3.75\,m, 4.75\,m, 5.75\,m]. This two-dimensional grid is repeated for three different heights (z-axis) to simulate where the receiver could be placed on the vehicle: car headlights (0.75\,m), dashboard (1\,m) or rear-view inside mirror (1.35\,m). 
The photodetector is connected to an oscilloscope to display and record the VLC signal coming from the traffic light. The oscilloscope is triggered by a sync output of the function generator  whose period, highlighted in Fig \ref{fig:signaltxrx}, corresponds to a full Manchester modulation cycle. The record length is 200 kpts at a sampling rate of 0.5 GHz, encompassing 10 full modulation periods. For each measurement we acquired both a single shot and the average of 4 traces.

\subsection{TX-RX Hardware Design}\label{sec:HW Design}

The \ac{vlc} TX and RX stages have been designed in order to feature analog bandwidths above 150 kHz and to reject low-frequency high-intensity  components coming from sunlight or traffic illumination without saturating the first amplification stage.
\subsubsection{Transmitter}\label{sec:TX}
\begin{figure}[tb]
\centering
\includegraphics[width=0.99\columnwidth, trim={0 0 0 0.5cm,clip}]{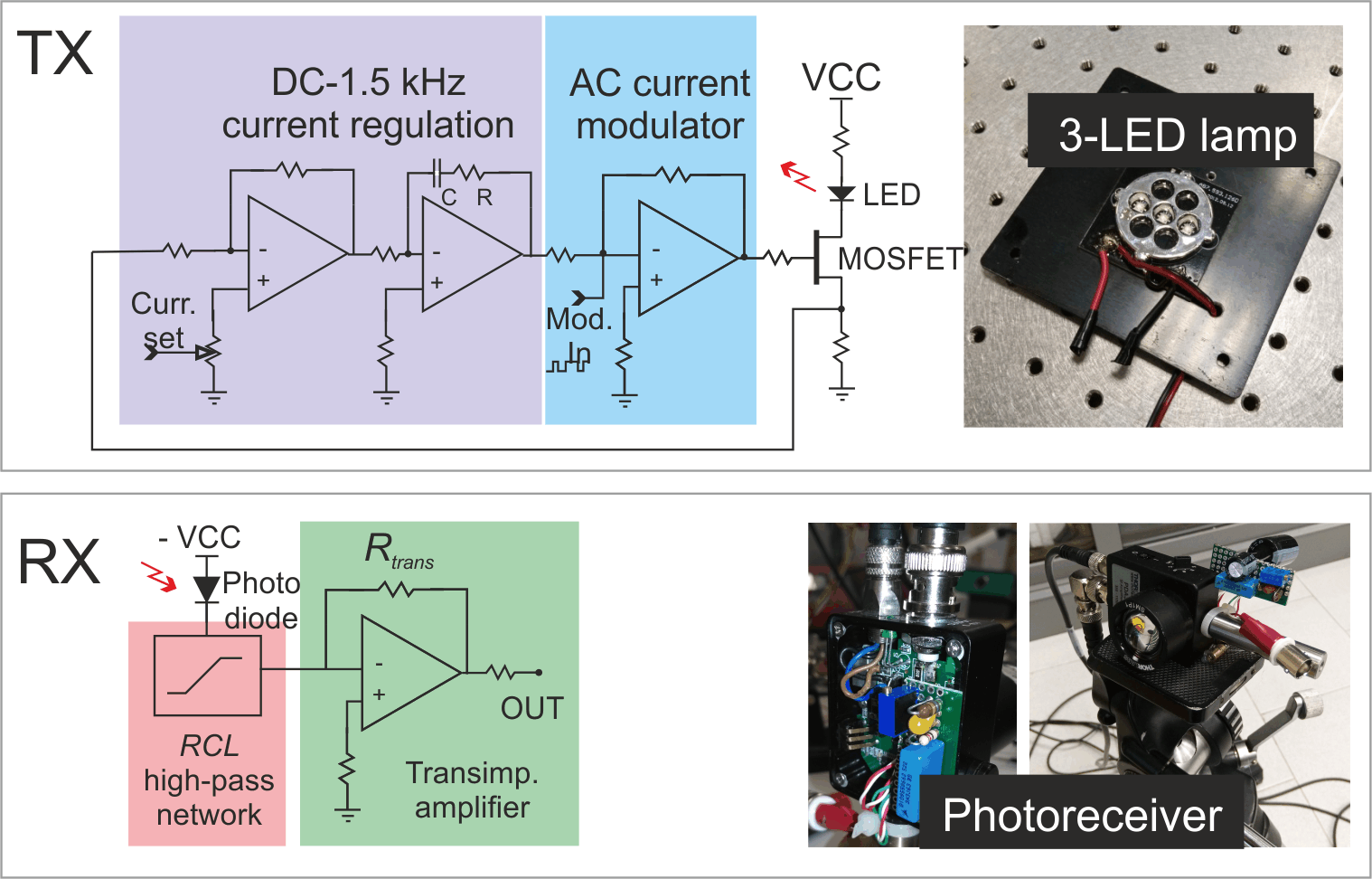}
\caption{Sketch of the TX-RX hardware. TX panel: A $PI$ regulation stage (purple area) stabilized the DC current value in the $RC$ bandwidth for LED to provide the required nominal light intensity, whereas a modulation stage (cyan area) inserts a high-frequency current modulation in the LED source (rightmost panel). RX panel: a Thorlabs PDA36-A is modified through the physical insertion of a high-pass RCL network among the photo-diode and the first stage of transimpendance amplification, to provide for beneficial AC decoupling of photo-current and avoid saturation due to stray ambient lights.}
\label{fig:TX_scheme}
\end{figure}

The schematic block diagram of TX hardware is reported in Fig.~\ref{fig:TX_scheme}. TX hardware is composed by the \acs{led} light source, and its current driver. Our implementation allows to generate the required current ($\simeq$ 0.7 A) for the traffic light \acs{led}s to provide the nominal luminous flux, as well as to insert a current modulation proportional to an external signal, which in turn allows for insertion of data streams into the optical carrier using any kind of protocol based on light intensity modulation. The modulator section (cyan shaded area of Fig.~\ref{fig:TX_scheme}) is placed after a \textit{P-I} regulation stage (purple shaded area), stabilizing the supply current, which is sensed by a precision resistor, from \ac{dc} to $\simeq$ 1.5 kHz, given by a proper adjustment of the servo \ac{rc} constant (see Fig.~\ref{fig:TX_scheme}). In such configuration, any modulation above the PI servo cut frequency $f_\mathrm{TX}= 1/2\pi RC\simeq$ 1.5 kHz will not be compensated by the \textit{P-I} loop, and will be added as a current modulation through the MOSFET transistor. The open loop bandwidth of the op-amp chain can virtually exceed several MHz, whereas, on the other hand, the large parasitic capacitance of the large-area LED module embedded in the traffic light lamp, as well as the presence of non linearity of the actuation chain in the open-loop response, affect the maximum achievable transmission bandwidth. We have limited the relative modulation amplitude to 30\% of the average \ac{dc} value of 700\,mA, in order to avoid a possible overburden of the LED sources due to excessive currents in the positive periods of the modulation pattern of the nominal \ac{dc} value. A function generator (Tektronix AFG1022) has been used to provide the modulation circuit with the modulation waveform (see Fig.~\ref{fig:signaltxrx}). The transmitted waveform has been chosen to match a \ac{ook} Manchester encoding, according  to  PHY  I  of  standard  IEEE  802.15.7  for  Outdoor VLC \cite{IEEE}. In order to embed all possible bit configurations of the Manchester encoding, the  data  package is constructed by attaching two square-wave blocks with frequencies of 50 kHz and 100 kHz, respectively, with a global periodicity of 40 $\mu$s, corresponding to the packet duration. The Manchester encoding grants a constant average signal (Fig.~\ref{fig:signaltxrx}), leading to a constant illumination intensity emitted by the traffic light lamp.

The (red) LED emitter (Lux Potentia OJ200-R07, 1A 12V) is composed a series of 3 high-power LEDs. In our scheme, the original power supply has been bypassed by our MOSFET-based current driver/modulator, while preserving the original case of the LED series, so that the global features of the traffic light illumination pattern are unaltered. \color{black}We also checked that different colour lamps do not have a significantly different frequency response to modulation, as the LED substrates have similar parasitic capacitance, so the bandwidth of the system is the same and the temporal dependence of Eq.\,\ref{eq.s(t)} is unaltered. \color{black} A red-coloured Fresnel lens shapes the beam according to the standards \cite{UNI11248,UNI13201-2} and increases the visibility at large distances. 

\subsubsection{Receiver}\label{sec:RX} 
The receiver hardware is implemented by modifying a Thorlabs PDA36-A active photodiode with a physical AC decoupling of the photodiode chip from the first transimpedance amplification stage. The time constant $f_\mathrm{RX}\simeq5$ kHz of such decoupling is realized through a multielement RLC network, with parameters chosen in such a way to filter out all of the unwanted low-frequency light variations (headlights, sunlight ecc.), still allowing for the modulation signal to pass through the first stage of the receiving electronics. Such configuration allowed for high RX gains without the risk of saturating the amplifier by a large light background, \color{black} and practically avoids the need for any coloured filter at receiver. \color{black} The photoreceiver gain has been chosen as the highest still preserving a bandwidth $>$ 150 kHz. The concentrator used in this RX setup (rightmost part of RX panel) is the shortest focal, low-cost, plastic lens we had available, i.e. a 25 mm diameter, $F$ = 30 mm focal length uncoated aspheric singlet. The choice of short-focal, aberration-corrected optical concentrators is crucial in order to keep a good trade-off between reasonably high acceptance angle and global optical gain when the solution to increase the input optics diameter beyond 1" is unaffordable. The data collection is accomplished by recording both the transmitted and received waveforms through a 70 MHz, 1Gs/s digital oscilloscope (Tektronix TBS2074).

\section{Propagation Model}
\label{propagation model}

The experimental (raw) data, recorded in each point of the 3D grid, has been processed to reduce the high-frequency noise effects and estimate the received amplitude in the frequency domain. The processing steps on the raw data can be summarized as follows:
\begin{itemize}
	\item Data Binning; 
	\item Fast Fourier Transform, normalized to a reference value chosen as maximum signal obtained in the ``reference point" (see Fig. \ref{fig:expsetup}); 
	\item Reconstruction of the amplitude of the received signal in each position $(x,y,z)$ of the grid. 
\end{itemize}
Each step is detailed in the following. 


We have observed that the frequency components of the received signal over 500\,kHz show a negligible amplitude, thus we applied a data binning, equivalent to a downsampling procedure, to reduce the frequency observation interval hence filtering out unnecessary high-frequency components, laying above our electronic system bandwidth. The binning procedure takes a cluster of consecutive (time) samples and replaces the cluster with the average value of the samples.  

The binning value can be defined as 
\begin{equation}
    \label{binning}
	N_{\text{bin}}=\dfrac{f_c}{2 f_{max}}
\end{equation}
	where $f_c$ is the initial sampling frequency of the received signal and $f_{max}$ is the highest frequency of the desired observation window. 
In our experiment, $N_{\text{bin}}=250$, since the sampling frequency of the oscilloscope is $250$\,MHz.

The ratio between the transfer function\footnote{The ratio $\dfrac{S_{\text{RX}}(f)}{S_{\text{TX}}(f)}$ is usually defined as the transfer function of a generic system (black box) that has $s_{\text{TX}}(t)$ as input and $s_{\text{RX}}(t)$ as output.} (TF) of the whole system calculated in each single point of the grid and the TF calculated in the reference position can be used to define the propagation model of the VLC signal from the traffic light to the vehicle
\begin{equation}\label{deltaH}
	\Delta H_{i}=\dfrac{H_{i}(f)}{H_{\text{ref}}(f)}
\end{equation}
with
\begin{equation}\label{deltaHs}
	H_{i}(f)=\dfrac{S_{\text{RX}_{i}}(f)}{S_{\text{TX}_{i}}(f)} ;\qquad  H_{\text{ref}}(f)=\dfrac{S_{\text{RX}_{\text{ref}}}(f)}{S_{\text{TX}_{\text{ref}}}(f)} 
\end{equation}
where $S_{\text{TX}_{i}}(f)$ and $S_{\text{RX}_{i}}(f)$ is the Fast Fourier Transform (FFT) of the transmitted and received signal, respectively, at location $i$, $H_{i}(f)$ represents the TF between the traffic light and the $i$th point on the grid and $H_{\text{ref}}(f)$ stands for the TF between the traffic light and the reference point. All measurements have been triggered on the same signal, and thus the FFT of transmitted signal will be the same for each point $S_{\text{TX}_{i}}(f)= S_{\text{TX}_{\text{ref}}}(f)$. Thus, Eq.~\eqref{deltaH} can be then rewritten as
\begin{equation}\label{pathloss}
	\Delta H_{i}=\dfrac{S_{\text{RX}_{i}}(f)}{S_{\text{RX}_{\text{ref}}}(f)}
\end{equation}
From Eq.\,\eqref{pathloss} we can deduce that $\Delta H_{i}$ depends only on the spectrum of the received signal at the $i$-th location and at the reference point\footnote{The effect of the thermal noise at the receiver has been neglected.}. In addition, we can safely assume that $\Delta H_{i}$ does not depend on the carrier and modulation frequencies, as the large spectral width of the optical carrier emitted by the LED source makes the contribution of absorption lines of air absolutely irrelevant for any channel loss, whilst the large difference between modulation and carrier frequencies avoids any frequency-dependent interference effect in line-of-sight tests. 

\begin{figure}[h]
	\centering
	\includegraphics[width=0.99\columnwidth, trim=0 2cm 0 0]{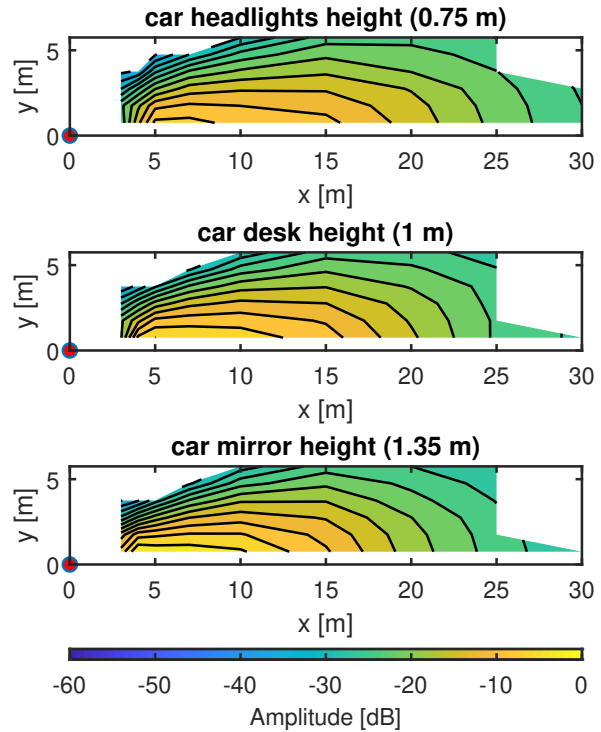}
	\caption{Amplitude map (dB) of the VLC signal $\Delta H_i$ measured over the grid for three different heights $[0.75, 1, 1.35]$\,m corresponding to car headlights, can desk and car internal mirror, respectively. The map reports the measurement data (intensity) over the field test grid shown in Fig.\,{\ref{fig:expsetup}}.}
	\label{fig:misure}
\end{figure}

Fig.\,\ref{fig:misure} shows the map of the amplitudes $\Delta H_{i}$ measured in each point of the grid at different height $[0.75, 1, 1.35]$\,m. Amplitudes are reported in logarithmic (dB) scale. We have used the point on the grid that shows the maximum amplitude (0 dB) as reference ("reference point" in Fig.\,\ref{fig:misure}). 
\color{black} The intensity is lower when the height is higher. In fact, at a distance $x=20$\,m the intensity clearly decreases moving from $h=0.75$\,m ($I \approx -15$\,dB) to $h=1.35$\,m ($I\approx -20$\,dB). This difference is then reflected in the accuracy map in Fig.\,\ref{fig:accuracy}.\color{black}

It is worth to point out that the ratio in Eq.~\eqref{pathloss} makes us to neglect the effects of the electronic components, $H_{\text{TX}}^{\text{el}}(f)$ and $H_{\text{RX}}^{\text{el}}(f)$, of the $i$-th received signal. We also note that spatial-dependent contributions of collecting optics can be neglected at the receiver, and hence the TF of the optical part at the receiver can be considered as constant for every point of the grid. This is due to the fact that the focused image of lamp is fully contained into the active area of the photodiode for all points in the measurement grid (see Sec.\,\ref{sec:systemModel}). 
Due to the operations described above, $\Delta H_i$ in Eq.\,\eqref{pathloss} does not depend on the frequency.

\subsection{Lambertian Model}
In the Lambertian model \cite{Akanegawa}, the path loss is given by
\begin{equation}\label{LittModel}
    I=\frac{I_{0}\,\, cos^{\nu}(\phi)}{d^2}
\end{equation}
where $I_0$ is the intensity in the axis of irradiance, $\phi$ is the irradiance angle, $\nu$ is a parameter for taking into account the influence of an optical lens in the LED emission, and $d$ is a distance between the transmitter and the receiver. \color{black}The angle $\phi$ is is referred to the optical axis of traffic light lens, identified as the line of maximum irradiance of the source (in our case, this is found to coincide with a line connecting the center of the  LED lamp and the point $(14,0,0)$\,m, i.e. located at 14\,m on the ground in front of the traffic light). Assuming rotational symmetry around the optical axis, the angle $\phi$ is illustrated in Fig.\,\ref{fig:angles}. \color{black}

\subsection{Proposed models for Intensity}
\begin{figure}[h]
	\centering
	\includegraphics[width=0.99\columnwidth]{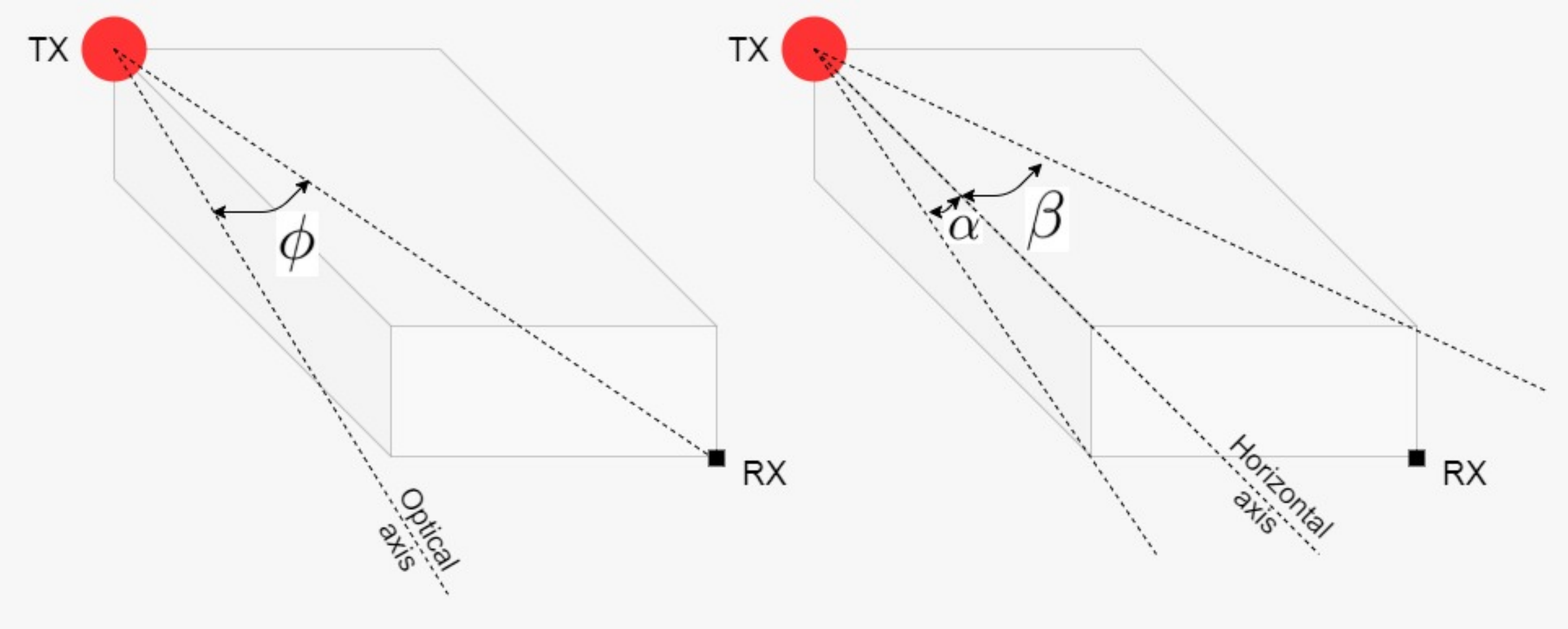}
	\caption{\color{black}[Left side] Angle $\phi$ of the Lambertian model \eqref{LittModel}. [Right side] Angles $\alpha$ and $\beta$ of the proposed models \eqref{modello1}, \eqref{modello2}. The horizontal axis is parallel to the road axis.\color{black}}
	\label{fig:angles}
\end{figure}

To mathematically characterize the propagation of the VLC signal, we aim to model the amplitude $\Delta H_i$ of the received signal over the grid, which in turn is quantified by its intensity. The intensity of the received signal depends on the position of the receiver in spherical coordinates 
\[
I(\alpha, \beta, d)
\]
where $\alpha$ is the elevation, $\beta$ is the azimuth and $d$ is the distance between the transmitter and the receiver. \color{black}Angles $\alpha$ and $\beta$ are shown in Fig.\,\ref{fig:angles}. 
\color{black}

The first model proposed here \eqref{modello1} is derived by modifying the well-known Lambertian propagation model \cite{Akanegawa}. Here the optical intensity is taken as: 
\begin{equation}
    I_1(\alpha,\beta,d)=\frac{f(\alpha,\beta)}{d^{2}}\,,
    \label{modello1}
\end{equation}
scaling as inversely proportional to the square of the distance. Differently from Lambert's law, the numerator has been generalized as a function of $\alpha$ and $\beta$.


The Lambert's cosine propagation law  assumes that the source of light is a homogeneous diffuser. Whilst remaining a valid approximation in the analysis of emission of many direct sources as LEDs, this hypothesis can be highly inaccurate in case of shaped beam patterns, i.e. in presence of lenses (as in traffic lights). Thus, the second propagation model proposed here only preserves a global $1/d^2$ dependence of radiated optical intensity:
\begin{equation}
    I_2(\alpha,\beta,d)=\frac{f(\alpha,\beta,d)}{d^{2}}\,.
    \label{modello2}
\end{equation}
The numerator is instead a generic polynomial function of ($\alpha$, $\beta$, $d$).

A more general approach can consider the optical intensity as a polynomial function of ($\alpha$, $\beta$, $d$) (with a given order) without an explicit $1/d^2$ weight, i.e.,  
\begin{equation}\label{eq.I3}
    I_3=f(\alpha,\beta,d)
\end{equation}

Depending on the specific needs and applications, one can choose a general or a more specific model which in general will feature very different convergence performances as a function of the number of parameters employed in the fitting procedure.

The multiple generalized linear regression (MGLR) method has been used to find the best fitting parameters for the three models $I_1$, $I_2$ and $I_3$.  Multiple linear regression is a generalization of simple linear regression to the case of more than one independent variable, and a special case of general linear models, restricted to one dependent variable. The generalized linear model (GLM) is typically used to model an irregular emission pattern \cite{Collett,Dobson,McCullagh} as it is a flexible generalization of ordinary linear regression that allows for response variables that have error distribution models other than a normal distribution. The GLM generalizes linear regression by allowing the linear model to be related to the response variable via a link function and by allowing the magnitude of the variance of each measurement to be a function of its predicted value. The output of the MGLR method is a set of polynomial coefficients of the function that fits the intensity.

In our procedure, the inputs of the MGLR method are the following parameters: the model under evaluation ($I_1$, $I_2$, $I_3$), the maximum order of the polynomial function for each variable ($\alpha$, $\beta$, $d$), the statistical distribution of the error (Normal, Poisson, Gamma)\footnote{Linear regression models describe a linear relationship between a response and one or more predictive terms. Many times, however, a nonlinear relationship exists. Nonlinear Regression describes general nonlinear models. A special class of nonlinear models, called generalized linear models, uses linear methods. Ordinary linear regression can be used to fit a straight line, or any function that is linear in its parameters, to data with normally distributed errors. This is the most commonly used regression model; however, it is not always a realistic one. Generalized linear models extend the linear model in two ways. First, assumption of linearity in the parameters is relaxed, by introducing the link function. Second, error distributions other than the normal can be modeled \cite{mccullagh1989generalized}.}

Let $y_1, \cdots, y_n$ denote $n$ independent observations on a response. We treat $y_i$ as a realization of a random variable $Y_i$. In the general linear model we assume that $Y_i$ has a normal distribution with mean $\mu_i$ and variance $\sigma$. 
We further assume that the expected value $\mu_i$ is a linear function of $p$ predictors that take values $\mathbf{x}_i=(x_{i1}, x_{i2}, \cdots, x_{ip})$ for the i-th case, so that
\begin{equation}\label{eq:GLM}
    E[Y_i]=\mu_i=\mathbf{x}_i^T \mathbf{b}
\end{equation}
where $\mathbf{b}$ is a vector of unknown parameters. Once found $\mathbf{b}$ from \eqref{eq:GLM} we can write 
\begin{equation}
    \mu_i= b_0 + b_1 x_{i1}+b_2 x_{i2}+ \cdots + b_p x_{ip}
\end{equation}

Different (from Normal) statistical distribution of the error can be used by taking into account the generalized linear model.
In a generalized linear model, the outcome $Y_i$ of the dependent variables is assumed to be generated from a particular distribution in the exponential family (Normal, Poisson and Gamma). The vector of the mean $\mu_i$ of the distribution depends on the independent variables $\mathbf{x}_i$ through
\begin{equation}
    \mu_i = g^{-1}\left( \mathbf{x}_i^T \mathbf{b} \right) 
\end{equation}
where $g(\cdot)$ is the \emph{link} function. 

To find the highest polynomial order for each of the variables ($\alpha, \beta, d$) to be inserted into the MGLR method still avoiding the overfitting problem, a k-fold method is used \cite{James,Devijver,Lachenbruch}.
The k-fold has been applied to every model ($I_1$, $I_2$, $I_3$) for each one of the error distribution (Normal, Poisson, Gamma). Thus, nine models have been evaluated. The polynomial order of $\alpha$ and $d$ ranges from 1 to 9, while the order of $\beta$ can assume only even values, from 2 to 8, because of azimuthal-symmetric nature of the emitted pattern. 

\color{black}Incidentally, by assuming that Fresnel lenses provide the same beam shape for the three LED lamps (red, orange and green), we remark that our model allows to directly retrieve intensity maps for all of them, as the only difference among the three is given by the different height with respect to ground and by the relative position of the emission wavelength peak with respect to the responsivity curve of Silicon photodiode. This can be simply taken into account with a different attenuation factor (depending on the specific colour) for all of the three models given by Eqs.\,\eqref{modello1}, \eqref{modello2} and \eqref{eq.I3}.\color{black}


\section{Results}\label{sec:Results}

\begin{figure}
    \centering
    \begin{subfigure}[t]{0.5\textwidth}
    \includegraphics[width=0.99\columnwidth]{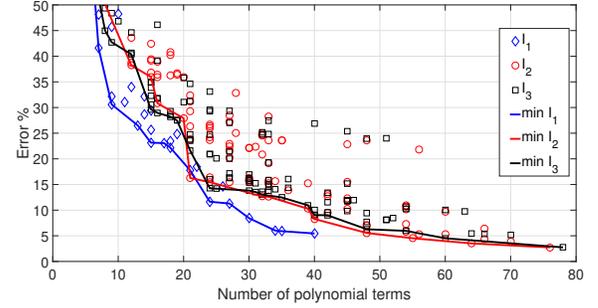}
    \caption{Normal distribution.}
    \label{fig:regLIN}
    \end{subfigure}
    \vfill
    \begin{subfigure}[t]{0.5\textwidth}
    \includegraphics[width=0.99\columnwidth]{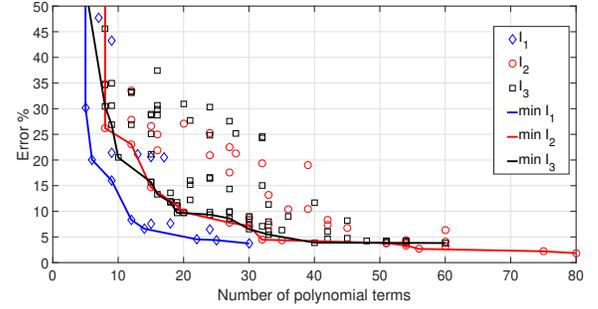}
    \caption{Gamma distribution.}
    \label{fig:regGAM}
    \end{subfigure}
    \begin{subfigure}[t]{0.5\textwidth}
    \includegraphics[width=0.99\columnwidth]{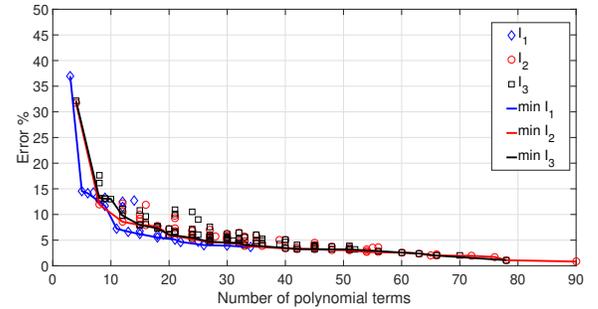}
    \caption{Poisson distribution.}
    \label{fig:regPOI}
    \end{subfigure}
    \caption{Percentage error for each of the proposed models $I_1(\alpha,\beta,d)$, $I_2(\alpha, \beta, d)$, $I_3(\alpha, \beta, d)$ as a function of the number of polynomial terms with a different distributions of the error in the MGLR method. Solid lines are a guide to the eye connecting the best parameter configuration minimizing the error for a specific number of terms.}
\end{figure}

The percentage error for each of the proposed models $I_1(\alpha,\beta,d)$, $I_2(\alpha, \beta, d)$, $I_3(\alpha, \beta, d)$ as a function of the number of polynomial terms with a Normal (Fig.~\ref{fig:regLIN}), Gamma (Fig.~\ref{fig:regGAM}) e Poisson (Fig.~\ref{fig:regPOI}) distribution of the error in the MGLR method. Each point corresponds to a specific parameter configuration. For example, one red square in Fig.~\ref{fig:regLIN} represents the error produced by the model $I_3(\alpha, \beta, d)$ with a specific number of polynomial terms for $\alpha$, $\beta$ and $d$ coming  out from the application of the k-fold and MGLR procedures. The error is calculated as the RMSE between the original measured data and the corresponding values of the polynomial fitting function over the whole grid. Solid lines are a guide to the eye connecting the best parameter configuration minimizing the error for a specific number of terms.
Given a specific error threshold, it is possible to reach it with the lowest number of polynomial terms, after which the minimization procedure looses efficiency. 

As it can be seen in Fig.~\ref{fig:regLIN}, the error decreases faster for model $I_1(\alpha,\beta,d)$. The other two models can get more accurate, at the expenses of introducing a high number of polynomial terms. Hence depending on the particular needs one can either choose a more ``efficient" or a more ``accurate" model for the VLC system.

If a Gamma or Poisson distribution is used in the MGLR method, the error decreases faster than using the Normal distribution. The $I_2$ model with Poisson distribution reaches the lowest error, but it requires a high number of polynomial terms (80) for the fitting function. If a low number of polynomial terms is desired, the model $I_1$ with Gamma or Poisson distribution reaches a lower error. A low number of polynomial terms is useful for extending the model to an area larger than the one where the measurements are taken. The k-fold method is used to avoid the overfitting problem in the area of the measurements. 

Only two models have been compared to define a propagation model which is valid for distances out of measurement's range. The first one is $I_1$ with Gamma distribution, while the second is $I_2$ with Poisson distribution. The first one has been selected because the error decreases more rapidly, the second one has been selected because it reaches the lowest error.


In case of $I_1(\alpha,\beta,d)$ with Gamma distribution, the vector $\V{y}$ of the intensity measured over the grid is
\begin{equation}\label{eq.y}
    \V{y} = (\Delta H_1 d_1^2, \Delta H_2 d_2^2, \cdots, \Delta H_n d_n^2)
\end{equation}
where $n$ is the number of points over the grid of the measurements, $\Delta H_i$ is the intensity (in dB) in the $i$-th point of the grid and $d_i$ is the distance of the $i$-th point of the grid from the source of the VLC signal (the traffic light ). 
The predictor matrix is composed by
\begin{equation}
    \M{X} = [\V{x}_1 \,\,\, \V{x}_2 \,\,\, \cdots \,\,\, \V{x}_n]^T
\end{equation}
with
\begin{equation}
    \V{x}_i = [1 \,\,\, \V{\alpha}_i \,\,\, \V{\beta}_i \,\,\, \V{\alpha}_i \otimes \V{\beta}_i]
\end{equation}
where $\otimes$ stands for the element-by-element product operator,  $\V{\alpha}_i=(\alpha_{i}, \alpha_i^2, \cdots, \alpha_{i}^p)$ is the vector of the elevation angles and $\V{\beta}_i=(\beta_{i}^2, \beta_i^4 \cdots, \beta_{i}^{2q})$ is the vector of the azimuth (squared) angles. The parameters $p$ and $2q$ represent the maximum polynomial order for the elevation parameter $\alpha$ and the azimuth parameter $\beta$, respectively. 

In case of $I_3(\alpha, \beta, d)$ with Poisson distribution, the vector $\V{y}$ of the intensity measured over the grid is the same \eqref{eq.y}, while 
\begin{equation}
     \V{x}_i = [1 \,\,\, \V{\alpha}_i \,\,\, \V{\beta}_i \,\,\, \V{d}_i \,\,\, \V{\alpha}_i \otimes \V{\beta}_i \otimes \V{d}_i]
\end{equation}
where $\V{d}_i=(d_{i}, d_i^2, \cdots, d_{i}^r)$. The parameter $r$ is the maximum polynomial order for the distance parameter $d$.

The objective is to find the coefficients $\V{b}=(b_1, \cdots, b_m)^T$ (with $m=pq+1$) so that
\begin{equation}\label{eq.modello4}
    E[\mathbf{y}]^{-1} = \mathbf{X} \mathbf{b}
\end{equation}
in case of model $I_1(\alpha,\beta,d)$ with Gamma distribution, and \begin{equation}\label{eq.modello8}
    \log\left(E[\mathbf{y}]\right) = \mathbf{X} \mathbf{b} 
\end{equation}
in case of model $I_2(\alpha, \beta, d)$ with Poisson distribution (with $m=pqr+1$). 

Table\,\ref{tab.modello4} and \ref{tab.modello8} show the different solutions for \eqref{eq.modello4} and \eqref{eq.modello8}, respectively, increasing the order of the polynomial function. The error associated to each model compared with the measurements is reported as average percentage error 
\begin{equation}
\epsilon_{\text{perc}}=\frac{1}{n} \sum_{i=1}^n \frac{|y_i-I_{li}|}{y_i} \times 100
\end{equation}
as well as root mean squared error (RMSE)
\begin{equation}
\epsilon_{\text{RMSE}} = \sqrt{\frac{\sum_{i=1}^n (y_i - I_{li})^2}{n}} 
\end{equation}
with $i=1, \cdots, n$ and $l=1,2,3$. 

\begin{table}[H]
\caption{Number of terms, orders of the variables and error for the solution \eqref{eq.modello4} of model $I_1(\alpha,\beta,d)$ with Gamma distribution. Only values of $(\alpha, \beta)$ that minimize the error are reported (see the curve min\,$I_1$ in Fig.\,{\ref{fig:regGAM}}).}\centering
\begin{tabular}{|c|c|c|c|c|} \hline 
 No. of terms & Order of $\alpha$  & Order for $\beta$ & Error (\%) & RMSE \\ \hline
 30 & 8 & 6 & 3.7307 & 0.014001 \\
 25 & 6 & 8 & 4.3831 &	0.017496   \\
 22 & 6 & 6 &  4.5449 &	0.018839 \\
 14 & 4 & 6 & 6.5972 &	0.02563 \\
 12	& 4	& 4	& 8.3422 &	0.030691 \\
 9 &	2	& 6	& 16.002 &	0.061841 \\
6 &	2 &	4 &	20.056 &	0.084346 \\
5 &	2 &	2 &	30.182 &	0.10461 \\ \hline 
\end{tabular}\label{tab.modello4}
\end{table}

\begin{table}[H]
\caption{Number of terms, orders of the variables and error for the solution \eqref{eq.modello8} of model $I_2(\alpha, \beta, d)$ with Poisson distribution. Only values of $(\alpha, \beta, d)$ that minimize the error are reported (see the curve min\,$I_2$ in Fig.\,{\ref{fig:regPOI}}).}\centering
\begin{tabular}{|c|c|c|c|c|c|} \hline 
 No. terms & Order $\alpha$  & Order $\beta$ & Order $d$ & Error (\%) & RMSE \\ \hline
 90 &	8 &	8 &	2 & 0.8221	& 0.0016 \\
 66	& 7	& 6	& 2	& 1.9923	& 0.0049   \\
60	& 6	& 4	& 4	& 2.5728	& 0.006811 \\
48	& 5	& 6	& 3	& 3.0709	& 0.010701 \\ 
40	& 6	& 6	& 1	& 3.3717	& 0.010796 \\ 
20	& 5	& 2	& 1	& 5.9458	& 0.014649 \\
12 & 3	& 2	& 1	& 8.5654	& 0.029661 \\
4 	& 1	& 2	& 1	& 31.776	& 0.056186 \\ \hline 
\end{tabular}\label{tab.modello8}
\end{table}

The polynomial function for the model $I_1(\alpha,\beta,d)$ \eqref{modello1} with Gamma distribution with 12 terms is 

\begin{equation}
\begin{split}
I_1(\alpha,\beta,d) &= \frac{1}{d^2} ( b_1+b_2\alpha+b_3\beta^2+b_4\alpha^2+b_5\alpha \beta^2+b_6\beta^4\\
    &+b_7\alpha^3+b_8\alpha^2\beta^2+b_9\alpha\beta^4+b_{10}\alpha^4+b_{11}\alpha^3\beta^2\\
    &+b_{12}\alpha^2\beta^4)^{-1}
    \end{split}
\end{equation}
while for the model $I_2(\alpha, \beta, d)$ \eqref{modello2} with Poisson distribution with 12 terms is
\begin{equation}
\begin{split}
    I_2(\alpha, \beta, d) & =  \frac{1}{d^2} \exp\{ b_1+b_2\alpha+b_3\beta^2+b_4d+b_5\alpha^2\\
    & +b_6\alpha\beta^2+b_7\alpha d+b_8\beta^2 d+b_9\alpha^3+b_{10}\alpha^2\beta^2\\
    & +b_{11}\alpha^2 d+b_{12}\alpha\beta^2 d\}
    \end{split}
\end{equation}
and the value of the coefficients is reported in Table\,\ref{tab.bcoeffmod4}. We selected the 12 terms since this choice provides in both models a fair tradeoff between performance (error below $10\%$) and complexity. 

It is important to note that the element $b_j$ of vector $\V{b}$ is set to zero if the corresponding order of the associated variables ($\alpha$, $\beta$, $d$) exceeds the maximum between the selected order of $\alpha$, $\beta$ or $d$. \textcolor{black}{The models are represented in spherical coordinates, so the effect of the heights is included.} 

\begin{table}[H]
\caption{Values of the coefficient $\mathbf{b}$ for the models $I_1(\alpha,\beta,d)$ with Gamma distribution and $I_2(\alpha, \beta, d)$ with Poisson distribution.}\centering
\begin{tabular}{|c|c|c|} \hline 
 Coefficient  & Value for $I_1$ model & Value for $I_2$ model \\ \hline
$b_1$ &	0.088395 & 6.1107 \\
$b_2$	& 1.8365 & 20.436\\
$b_3$	& 0.53823 & -9.8384\\
$b_4$	& 14.718 & -0.09868\\
$b_5$	& 6.3874 &  47.629\\
$b_6$	&0.92338 &-12.142\\
$b_7$	&46.406 &-1.0858\\
$b_8$	&26.178 &-0.08044\\
$b_9$	&-7.8413 &52.16\\
$b_{10}$	&52.665 &13.944\\
$b_{11}$	&39.219 &-1.4756\\
$b_{12}$	&-1.3364 & 0.90893 \\ \hline 
\end{tabular}\label{tab.bcoeffmod4}
\end{table}

\begin{figure}[ht]
    \centering
    \begin{subfigure}[t]{0.45\textwidth}
        \includegraphics[width=0.99\columnwidth]{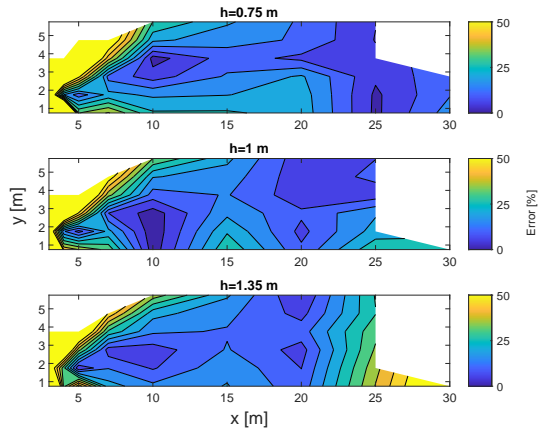}
    \caption{Accuracy of the conventional Lambertian model.}
    \label{fig:lambertian}
    \end{subfigure}
        \begin{subfigure}[t]{0.45\textwidth}
        \includegraphics[width=0.99\columnwidth]{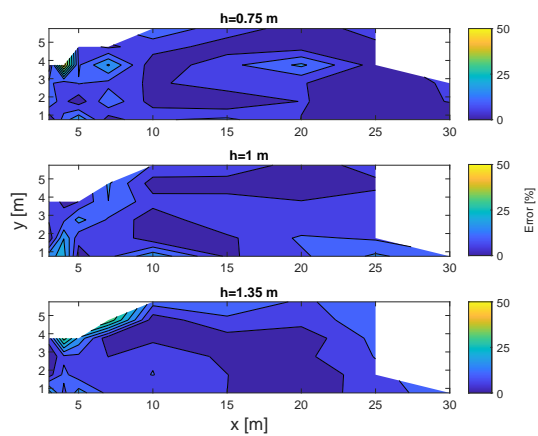}
    \caption{Accuracy of the $I_1$ model.}
    \label{fig:gamma}
    \end{subfigure}
    \begin{subfigure}[t]{0.45\textwidth}
        \includegraphics[width=0.99\columnwidth]{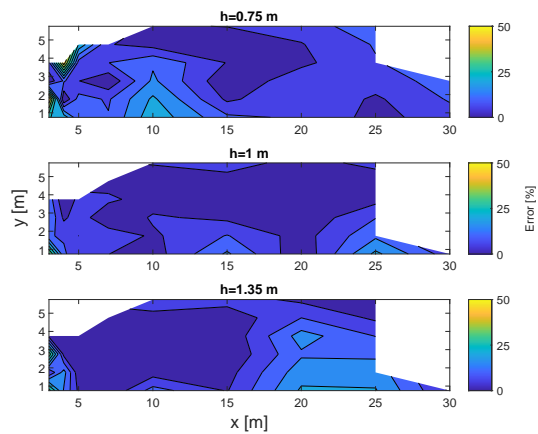}
    \caption{Accuracy of the $I_2$ model.}
    \label{fig:poisson}
    \end{subfigure}
    \caption{Comparison of the propagation models $I_1$ and $I_2$ with the conventional Lambertian model.}\label{fig:accuracy}
\end{figure}

The proposed models have been compared to the conventional Lambertian model {\cite{KomineNakagawa}}, widely used in literature. Fig.\,{\ref{fig:accuracy}} shows the accuracy of the Lambertian model as well as of the proposed new models, over the experimental measurements grid. It is important to note that the accuracy has been calculated only over the points where we had the value corresponding to a real measurement. We could not measure the intensity on the other side of the traffic-light since there was no lane available, as in a typical urban road.

The accuracy is calculated as the difference between the measured intensity in a point of the grid and the corresponding intensity of the fitting polynomial function of the model. 

As it can be seen the conventional model shows a significant higher error compared to the models proposed in this paper. The average RMSE of the conventional Lambertian is 0.097, while the one of the two proposed models (with, e.g., 12 parameters) is 0.030 and 0.029 (see Table\,{\ref{tab.modello4}} and {\ref{tab.modello8}}).


\begin{figure*}[h!]
    \centering
    \begin{subfigure}[b]{0.40\textwidth}
        \includegraphics[width=0.99\columnwidth]{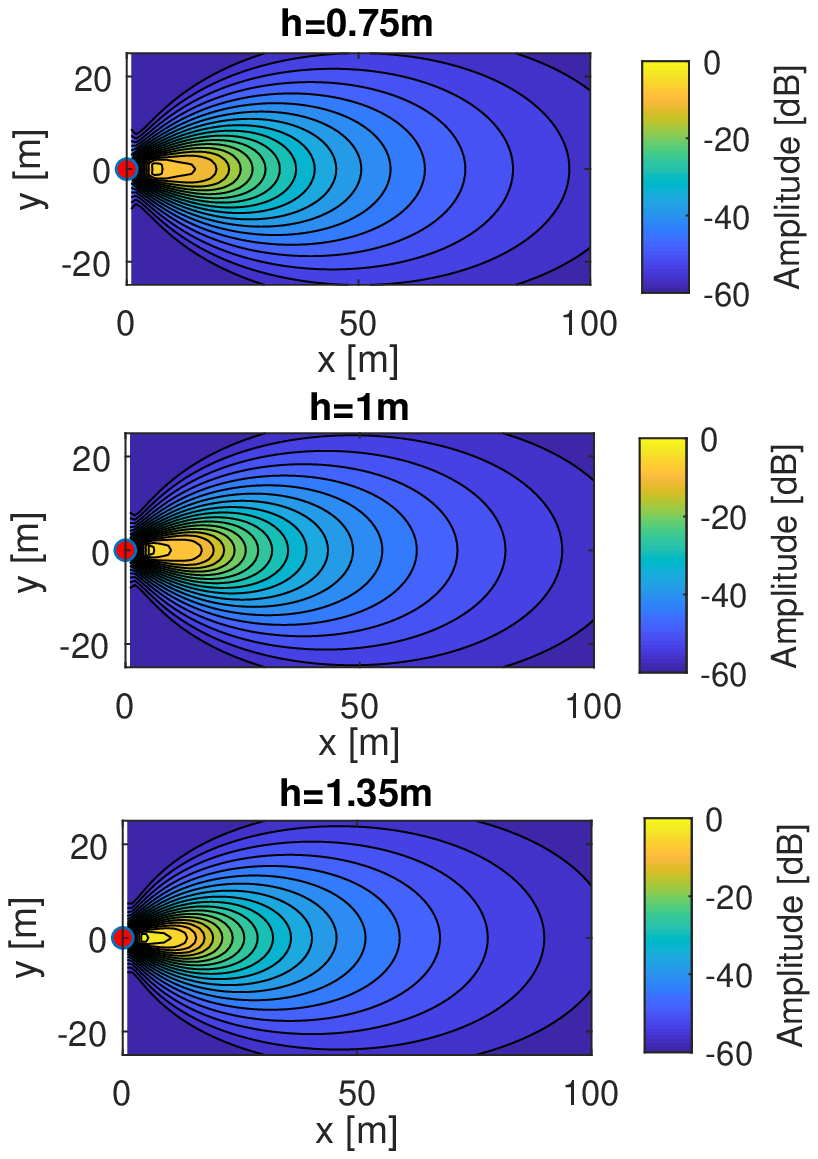}
    \caption{Intensity of the propagation model $I_1(\alpha,\beta,d)$ with Gamma distribution in an area $100 \times 50$\,m. The photodetector heights are $\{0.75, 1, 1.35\}$\,m. The polynomial order for $\alpha$ an $\beta$ is 4.}
    \label{fig:LGA44}
    \end{subfigure}
    \begin{subfigure}[b]{0.40\textwidth}
        \includegraphics[width=0.99\columnwidth]{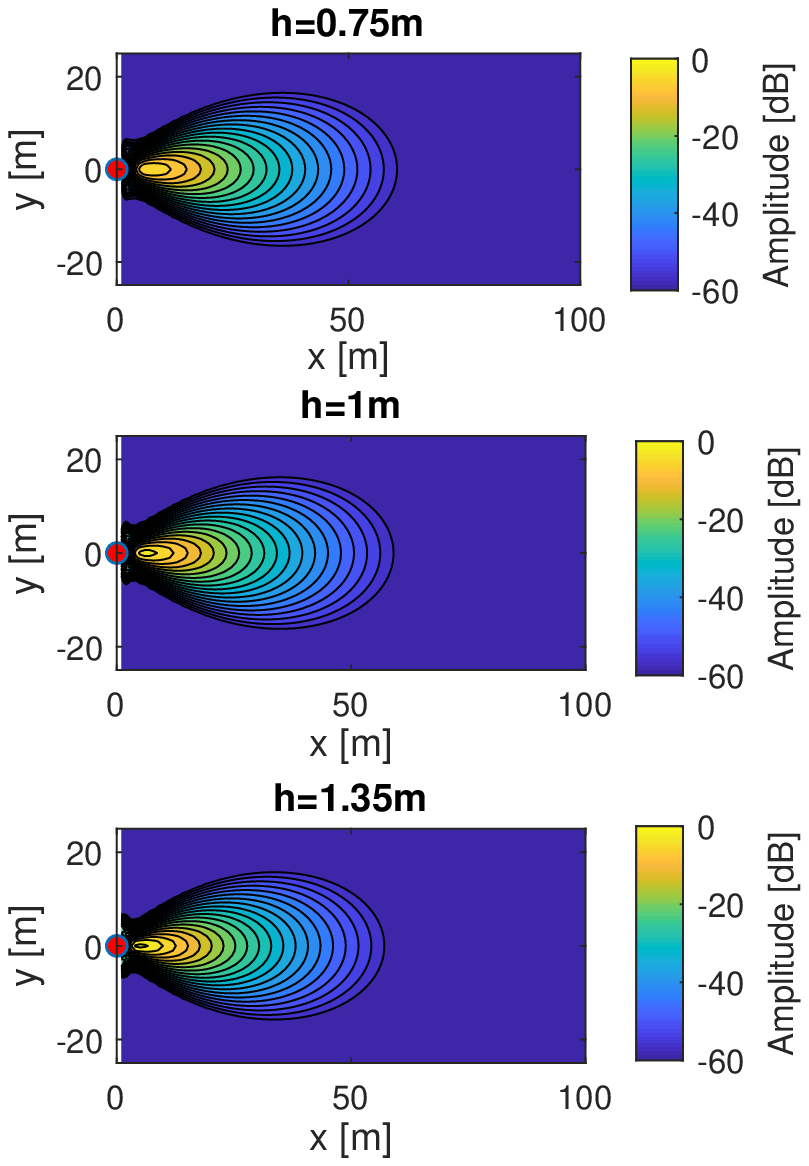}
    \caption{Intensity of the propagation model $I_2(\alpha, \beta, d)$ with Poisson distribution in an area $100 \times 50$\,m. The photodetector heights are $\{0.75, 1, 1.35\}$\,m. The polynomial order for $\alpha$ is 3, for $\beta$ is 2 and for $d$ is 1.}
    \label{fig:IPO321}
    \end{subfigure}
    \begin{subfigure}[b]{0.50\textwidth}
        \includegraphics[width=0.99\columnwidth]{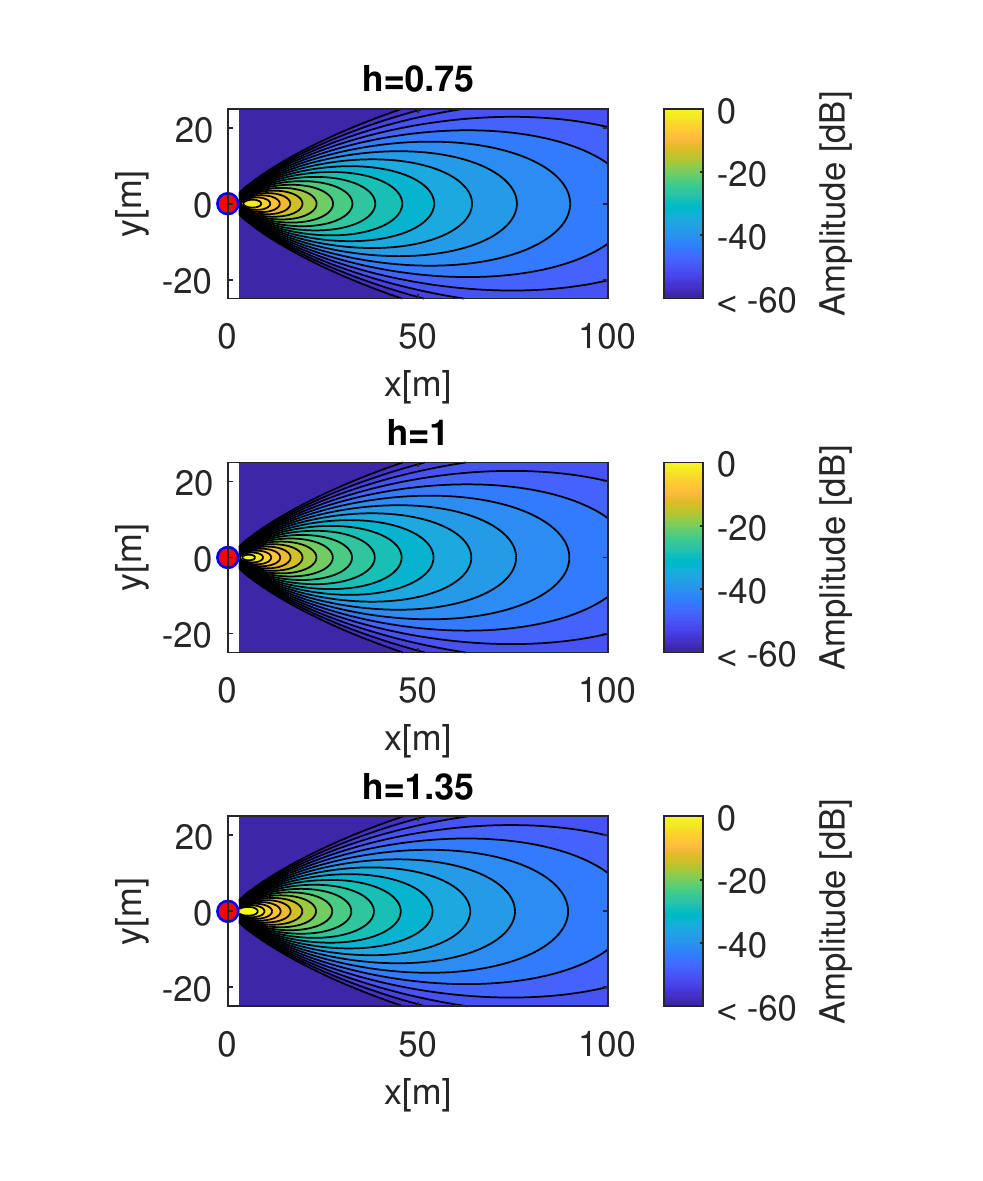}
    \caption{Intensity of the Lambertian model in an area $100 \times 50$\,m. The photodetector heights are $\{0.75, 1, 1.35\}$\,m.}
    \label{fig:555}
    \end{subfigure}
    \caption{Comparison of the propagation models $I_1$ and $I_2$ with the conventional Lambertian model.}\label{fig:grafEST}
\end{figure*}

Fig.~{\ref{fig:grafEST}} shows the intensity of the propagation models $I_1(\alpha,\beta,d)$ model with Gamma distribution and $I_2(\alpha, \beta, d)$ with Poisson distribution.
Figs.~{\ref{fig:LGA44}} and {\ref{fig:IPO321}} report the  intensity of the propagation models described above, extended to an area ($100 \times 50$\,m) larger than the measurements grid.

The two models shown in Fig.\,{\ref{fig:grafEST}} have the same complexity (12 terms) and the same accuracy (see Table I and II). The accuracy panels (right part of Fig.\,{\ref{fig:grafEST}}) show local deviations from data not exceeding 20\%. These accuracy maps allow for eventual selection of one model depending on distance and/or height of the receiver. For example, in case of receiver placed at a distance of 20\,m and height of 1.35\,m model $I_1$ is more accurate than model $I_2$ as depicted in Fig.\,{\ref{fig:grafEST}}. The behaviour of the accuracy seems not to have a direct dependence on the distance, but this is rather dominated by numerical fluctuations of the polynomial functions involved.
\color{black}
The computational cost to obtain the Lambertian model is $\mathcal{O}$($\nu$+3), where $\nu$ is defined in Eq.\,\eqref{LittModel}. It is important to note that the index $\nu$ is usually high. In our case, the value of $\nu$ that minimizes the error of the Lambertian model is $\approx 22$. Concerning the two proposed models, Eq.\,\eqref{eq.modello4}, which derives the polynomial coefficients of the model $I_1$, and Eq.\,\eqref{eq.modello8}, which derives the model $I_2$, show a computational cost of $\mathcal{O}$($m^5$). In the paper we have evaluated the two models with $m = 12$ coefficients, since this assured an average error below 10\%. The accuracy-complexity tradeoff of the proposed models should be selected based on the application scenario. Assume to have an urban scenario with, e.g., tens/hundreds of traffic-lights in a specific perimeter of a modern city. In order to have a global model of propagations in a reasonable time, we could choose to limit the number of parameters in the regression model, although this implies a lower accuracy. 
\color{black}

Importantly, although we have not measured directly the bit error rate (BER) of the VLC communication link, the performance can be analytically derived from the intensity measurements. The results show that a probability of error of $10^{-3}$ can be provided up to 30\,m away from the source. Mathematical details are shown in Appendix~\ref{Performance Analysis}.

\section{Conclusion}\label{sec:conclusion}
This paper presents an extensive measurement campaign aimed at VLC channel characterization for I2V communication, carried out by using a real traffic-light in a typical urban road. 
The receiver has been located at three different heights corresponding to car headlights, dashboard and internal mirror, respectively.
A specific hardware for modulating the LED of the traffic light has been designed and implemented. The data have been then used to mathematically model the transmission patter and the propagation channel. Three models have been proposed and compared in terms of complexity and accuracy, together with the conventional Lambertian model. The results show that the Lambertian model is less accurate in describing the \ac{vlc} transmission-propagation from traffic-light to vehicles when real lamps are considered. The proposed models provide for a a more accurate description of the transmission channel, and can be extended to distances wider than those considered during the measurements campaign.
\textcolor{black}{In particular, the Lambertian model provides an average error of 25\%, while the two proposed polynomial models obtain 8\% with 12 coefficients. Anyway, the regression process to calculate the coefficients of the two proposed polynomial models has a computational cost higher than the Lambertian model.}

The models can be reliably used to evaluate the performance of VLC technology for automotive applications. In addition, the theoretical error probability of the VLC signal has been derived by using the experimental data. The results highlight that an uncoded error probability of $10^{-3}$ is achievable at 30\,m distance, showing that our implementation can be safely used for data service in I2V applications. \color{black} Furthermore, our work can be a valuable starting point to assess the impact of more generic weather scenarios on I2V VLC links, such as those in presence of fog and/or rain. \color{black}

\section*{Acknowledgment}
This work has been performed by the Visible Light Communications Research laboratory (VisiCoRe, https://visicore.dinfo.unifi.it/), a partnership of University of Florence, CNR-INO and LENS.   

Authors would like to thank the company ILES srl (http://www.ilessrl.com) for supplying the traffic-light and for the important support given during the measurements campaign.

\bibliographystyle{IEEEtran}
\bibliography{refs}

\begin{thebibliography}{10}
\providecommand{\url}[1]{#1}
\csname url@samestyle\endcsname
\providecommand{\newblock}{\relax}
\providecommand{\bibinfo}[2]{#2}
\providecommand{\BIBentrySTDinterwordspacing}{\spaceskip=0pt\relax}
\providecommand{\BIBentryALTinterwordstretchfactor}{4}
\providecommand{\BIBentryALTinterwordspacing}{\spaceskip=\fontdimen2\font plus
\BIBentryALTinterwordstretchfactor\fontdimen3\font minus
  \fontdimen4\font\relax}
\providecommand{\BIBforeignlanguage}[2]{{%
\expandafter\ifx\csname l@#1\endcsname\relax
\typeout{** WARNING: IEEEtran.bst: No hyphenation pattern has been}%
\typeout{** loaded for the language `#1'. Using the pattern for}%
\typeout{** the default language instead.}%
\else
\language=\csname l@#1\endcsname
\fi
#2}}
\providecommand{\BIBdecl}{\relax}
\BIBdecl

\bibitem{KomineNakagawa}
T.~Komine and M.~Nakagawa, ``Fundamental analysis for visible-light
  communication system using {LED} lights,'' \emph{IEEE Transactions on
  Consumer Electronics}, vol.~50, no.~1, pp. 100--107, 2004.

\bibitem{Alfattani}
S.~Alfattani, ``Review of {LiFi} technology and its future applications,''
  \emph{Journal of Optical Communications}, 2018.

\bibitem{ChiHaas}
N.~Chi, H.~Haas, M.~Kavehrad, T.~D.~C. Little, and X.~Huang, ``Visible light
  communications: demand factors, benefits and opportunities,'' \emph{IEEE
  Wireless Communications}, vol.~22, no.~2, pp. 5--7, 2015.

\bibitem{Martinez}
F.~Martinez, C.~Toh, J.~C. Cano, C.~Calafate, and P.~Manzoni, ``Emergency
  services in future intelligent transportation systems based on vehicular
  communication networks,'' \emph{IEEE Intelligent Transportation Systems
  Magazine}, vol.~2, pp. 6--20, 2010.

\bibitem{Karagiannis}
G.~Karagiannis, O.~Altintas, E.~Ekici, G.~Heijenk, B.~Jarupan, K.~Lin, and
  T.~Weil, ``Vehicular networking: A survey and tutorial on requirements,
  architectures, challenges, standards and solutions,'' \emph{IEEE
  communications surveys and tutorials}, vol. accepted f, no.~4, pp. 584--616,
  2011.

\bibitem{WuWangYoun}
S.~Wu, H.~Wang, and C.~Youn, ``Visible light communications for 5{G} wireless
  networking systems: from fixed to mobile communications,'' \emph{IEEE
  Network}, vol.~28, no.~6, pp. 41--45, 2014.

\bibitem{Chi}
N.~Chi, \emph{Led-based Visible Light Communications}.\hskip 1em plus 0.5em
  minus 0.4em\relax Springer-verlag, Berlin AN, 2018.

\bibitem{Akanegawa}
M.~Akanegawa, Y.~Tanaka, and M.~Nakagawa, ``Basic study on traffic information
  system using {LED} traffic lights,'' \emph{IEEE Intelligent Transportation
  Systems}, vol.~2, no.~4, pp. 197--203, 2001.

\bibitem{Correa}
A.~Correa, A.~Hamid, and E.~Sparks, ``{Li-Fi} based smart traffic network,'' in
  \emph{IEEE Transportation Electrification Conference and Expo (ITEC)}, Long
  Beach, CA, USA, 2018, pp. 217--219.

\bibitem{BobanKousaridas}
M.~Boban, A.~Kousaridas, K.~Manolakis, J.~Eichinger, and W.~Xu, ``Connected
  roads of the future: Use cases, requirements, and design considerations for
  vehicle-to-everything communications,'' \emph{IEEE Vehicular Technology
  Magazine}, vol.~13, no.~3, pp. 110--123, 2018.

\bibitem{Hu}
F.~Hu, \emph{Vehicle-to-vehicle and Vehicle-to-infrastructure Communications: A
  Technical Approach}.\hskip 1em plus 0.5em minus 0.4em\relax CRC Press, 2018.

\bibitem{BobanManolakis}
M.~Boban, K.~Manolakis, M.~Ibrahim, S.~Bazzi, and W.~Xu, ``Design aspects for
  {5G} {V2X} physical layer,'' in \emph{IEEE Conference on Standards for
  Communications and Networking (CSCN)}, 2016, pp. 1--7.

\bibitem{LeeKwon}
S.~J. Lee, S.~Y. Kwon, S.~Y. Jung, and Y.~H. Kwon, ``Simulation modeling of
  visible light communication channel for automotive applications,'' in
  \emph{15th International IEEE Conference on Intelligent Transportation
  Systems}, Anchorage, Alaska, USA, 2012, pp. 1--7.

\bibitem{CuiChen}
K.~Cui, G.~Chen, Z.~Xu, and R.~D. Roberts, ``Experimental characterization of
  traffic light to vehicle {VLC} link performance,'' in \emph{2011 IEEE
  GLOBECOM Workshops (GC Wkshps)}, 2011, pp. 808--812.

\bibitem{Barry}
J.~R. Barry, J.~M. Kahn, W.~J. Krause, E.~A. Lee, and D.~G. Messerschmitt,
  ``Simulation of multipath impulse response for indoor wireless optical
  channels,'' \emph{IEEE Journal on Selected Areas in Communications}, vol.~11,
  no.~3, pp. 367--379, 1993.

\bibitem{LeePark}
K.~Lee, H.~Park, and J.~R. Barry, ``Indoor channel characteristics for visible
  light communications,'' \emph{IEEE Communications Letters}, vol.~15, no.~2,
  pp. 217--219, 2011.

\bibitem{Yaqoob}
I.~Yaqoob, I.~A.~T. Hashem, Y.~Mehmood, A.~Gani, S.~Mokhtar, and S.~Guizani,
  ``Enabling communication technologies for smart cities,'' \emph{IEEE
  Communications Magazine}, vol.~55, no.~1, pp. 112--120, 2017.

\bibitem{Ayub}
S.~Ayub, S.~Kariyawasam, M.~Honary, and B.~Honary, ``A practical approach of
  {VLC} architecture for smart city,'' in \emph{Loughborough Antennas and
  Propagation Conference (LAPC)}, Loughborough, 2013, pp. 106--111.

\bibitem{Nawaz_2019_IEEE}
T.~{Nawaz}, M.~{Seminara}, S.~{Caputo}, L.~{Mucchi}, F.~S. {Cataliotti}, and
  J.~{Catani}, ``{IEEE} 802.15.7-compliant ultra-low latency relaying {VLC}
  system for safety-critical {ITS},'' \emph{IEEE Transactions on Vehicular
  Technology}, vol.~68, no.~12, pp. 12\,040--12\,051, Dec 2019.

\bibitem{8864129}
S.~{Avătămăniţei}, A.~{Căilean}, E.~{Zadobrischi}, A.~{Done}, M.~{Dimian},
  and V.~{Popa}, ``\textcolor{black}{Intensive Testing of
  Infrastructure-to-Vehicle Visible Light Communications in Real Outdoor
  Scenario: Evaluation of a 50 meters link in Direct Sun Exposure},'' in
  \emph{2019 Global LIFI Congress (GLC)}, June 2019, pp. 1--5.

\bibitem{Moreira}
A.~J. Moreira, R.~T. Valadas, and A.~Duarte, ``Optical interference produced by
  artificial light,'' \emph{Wireless Networks}, vol.~3, pp. 131--140, 1997.

\bibitem{Islim}
M.~S. Islim, S.~Videv, M.~Safari, E.~Xie, J.~J.~D. McKendry, E.~Gu, M.~Dawson,
  and H.~Haas, ``The impact of solar irradiance on visible light
  communications,'' \emph{Journal of Lightwave Technology}, pp. 1--1, 2018.

\bibitem{Dahri}
F.~A. Dahri, S.~Ali, and M.~M. Jawaid, ``A review of modulation schemes for
  visible light communication,'' \emph{IJCSNS International Journal of Computer
  Science and Network Security}, vol.~18, no.~2, pp. 117--125, 2018.

\bibitem{UNI11248}
``{UNI 11248} - illuminazione stradale, selezione delle categorie
  illuminotecniche,'' Ente nazionale italiano di unificazione, Standard, 2016.

\bibitem{UNI13201-2}
``{UNI 13201-2} - illuminazione stradale, requisiti prestazionali,'' Ente
  nazionale italiano di unificazione, Standard, 2016.

\bibitem{IEEE}
\emph{Visible Light Communication}, IEEE Std. 802.15.7 WPAN Task Group 7 (TG7),
  http://www.ieee802.org/15/pub/TG7.html.

\bibitem{Collett}
D.~Collett, \emph{Modeling Binary Data}.\hskip 1em plus 0.5em minus 0.4em\relax
  New York: Chapman and Hall, 2002.

\bibitem{Dobson}
A.~J. Dobson, \emph{An Introduction to Generalized Linear Models}.\hskip 1em
  plus 0.5em minus 0.4em\relax New York: Chapman and Hall, 1990.

\bibitem{McCullagh}
P.~McCullagh and J.~A. Nelder, \emph{Generalized Linear Models}.\hskip 1em plus
  0.5em minus 0.4em\relax New York: Chapman and Hall, 1990.

\bibitem{mccullagh1989generalized}
\BIBentryALTinterwordspacing
P.~McCullagh and J.~Nelder, \emph{Generalized Linear Models, Second Edition},
  ser. Chapman and Hall/CRC Monographs on Statistics and Applied Probability
  Series.\hskip 1em plus 0.5em minus 0.4em\relax Chapman \& Hall, 1989.
  [Online]. Available: \url{http://books.google.com/books?id=h9kFH2\_FfBkC}
\BIBentrySTDinterwordspacing

\bibitem{James}
G.~James, H.~T. Witten, D., and R.~Tibshirani, \emph{An Introduction to
  Statistical Learning}.\hskip 1em plus 0.5em minus 0.4em\relax Springer, 2013.

\bibitem{Devijver}
P.~A. Devijver and J.~Kittler, \emph{Pattern Recognition: A Statistical
  Approach}.\hskip 1em plus 0.5em minus 0.4em\relax London, GB: Prentice-Hall,
  1982.

\bibitem{Lachenbruch}
P.~A. Lachenbruch and M.~R. Mickey, ``Estimation of error rates in discriminant
  analysis,'' \emph{Technometrics}, vol.~10, no.~1, pp. 1--12, 1968.

\end{thebibliography}

\appendix

\subsection{Performance analysis}\label{Performance Analysis}
To evaluate the performance in terms of probability of error and maximum achievable bit-rate, we first estimate the noise level in each point of the measurements grid. The error probability of digital signalling in wireless channels is given by
\begin{equation}
    P_e \leq (S-1)Q\left( \sqrt{\frac{a^2\rho_{min}^2}{4N_0}} \right)
\end{equation}
where $S$ is the number of symbols in the digital constellation, $a$ is the fading coefficient, $\rho_{min}$ is the minimum distance between the symbols and $N_0$ is the noise spectrum density power. 

\begin{figure}[ht]
	\centering
	\includegraphics[width=0.99\columnwidth]{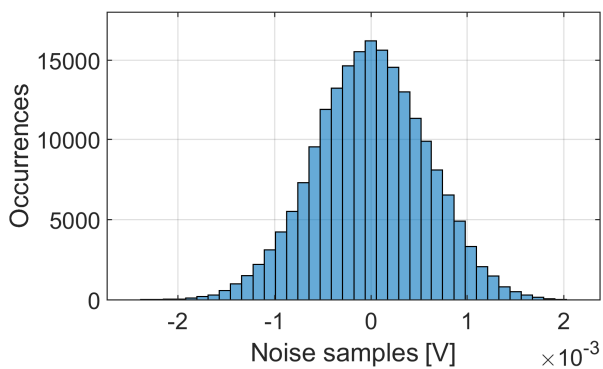}
    \caption{Histogram of the noise vector at the grid point $i$ is $(x=30, y=1.75, z=0.75)$\,m.}
	\label{fig:noise}
\end{figure}

Supposing than an AWGN model holds, the relation between the probability of error and the noise level becomes 
\begin{equation}
    P_e = Q\left( \sqrt{\frac{\rho_{min}^2}{2N_0}} \right)
\end{equation}
The distance $\rho_{min}$ depends on the specific constellation that has been transmitted. In our experiments, an antipodal (BPSK) constellation was used, thus the $P_e$ can be written as 
\begin{equation}\label{eq.PeSNR}
    P_e = Q\left( \sqrt{\frac{P_R}{\sigma_N^2}}\right)= Q\left( \sqrt{SNR} \right)
\end{equation}
where $P_R$ is the received power and $\sigma_N^2$ is the noise variance. 
To estimate the noise variance the following procedure has been carried out. Let us first remind that the received signal vector at location $i$ is 
\begin{equation}
    \V{s}_i = [\V{s}_{i1} \,\,\, \cdots \,\,\, \V{s}_{iN_b}]
\end{equation}
where $\V{s}_{ij}$ is the vector coming from the sampling of the waveform in \eqref{eq.waveform}. 
The received vector at the reference location ($i=\text{ref}$) is averaged over the repetition periods
\begin{equation}
    \V{\overline{s}}_{\text{ref}}= \frac{1}{N_b} \sum_{j=1}^{N_b} \V{s}_{\text{ref}\, j}
\end{equation}
Now, the estimated noise vector at location $i$ 
\begin{equation}\label{eq.noise}
    \V{w}_i = [\V{w}_{i1} \,\,\, \cdots \,\,\, \V{w}_{iN_b}]
\end{equation}
is calculated as the difference between the received signal vector at location $i$ and the averaged vector at reference location multiplied by $\Delta H_i$ \eqref{deltaH} 
\begin{equation}\label{eq.noise2}
  \V{w}_{ij}= \V{s}_{ij}-(\V{\overline{s}}_{\text{ref}} \Delta H_i)  \,\,\,\,\,\,\, j=1, \cdots, N_b
\end{equation}
The noise vector in \eqref{eq.noise} is then used to estimate the noise variance $\sigma_{w_i}^2$ at location $i$. An example of histogram of the noise vector $\V{w}_i$ is reported in Fig.\,\ref{fig:noise}, where the grid point $i$ is $(x=30, y=1.75, z=0.75)$\,m.
Fig.\,\ref{fig:noise} shows the occurrences (y-axis) of the noise samples (x-axis) as in \eqref{eq.noise2}. 

\begin{figure}[ht]
    \centering
    \includegraphics[width=0.99\columnwidth]{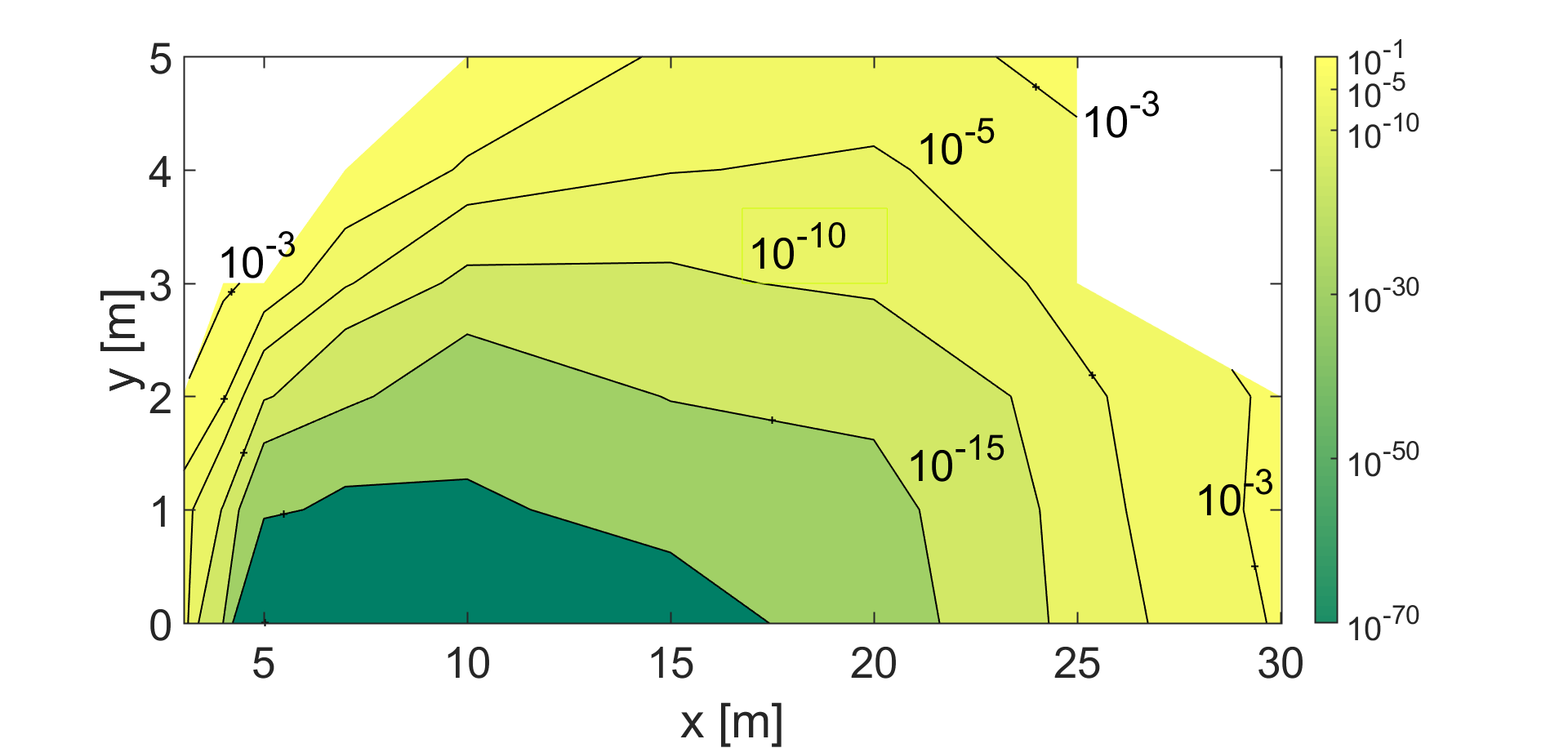}
    \caption{Map of the error probability over the measurements grid.}
    \label{fig:Pe}
\end{figure}

The signal-to-noise ratio (SNR) can thus be calculated as
\begin{equation}
    SNR_i = \frac{E[\V{\overline{s}}^2_{\text{ref}}]\Delta H_i^2}{\sigma_{w_i}^2}
\end{equation}
This result can be used to calculate the error probability \eqref{eq.PeSNR} in every point of the measurement grid. The map of the error probability is reported in Fig.\,\ref{fig:Pe} for the height $0.75$\,m.  
A probability of error of $10^{-3}$ (uncoded) can be provided even at $30$\,m away from the traffic light. The map shows the error probability over the perimeter where the experimental measurements have been carried on. The intensity models proposed in this paper can be anyway used to calculate the performance of a VLC system in different perimeters.

\subsection{FOV and reflections form tarmac}\label{app:fov}
\begin{figure}[htb]
    \centering
    \includegraphics[width=0.99\columnwidth]{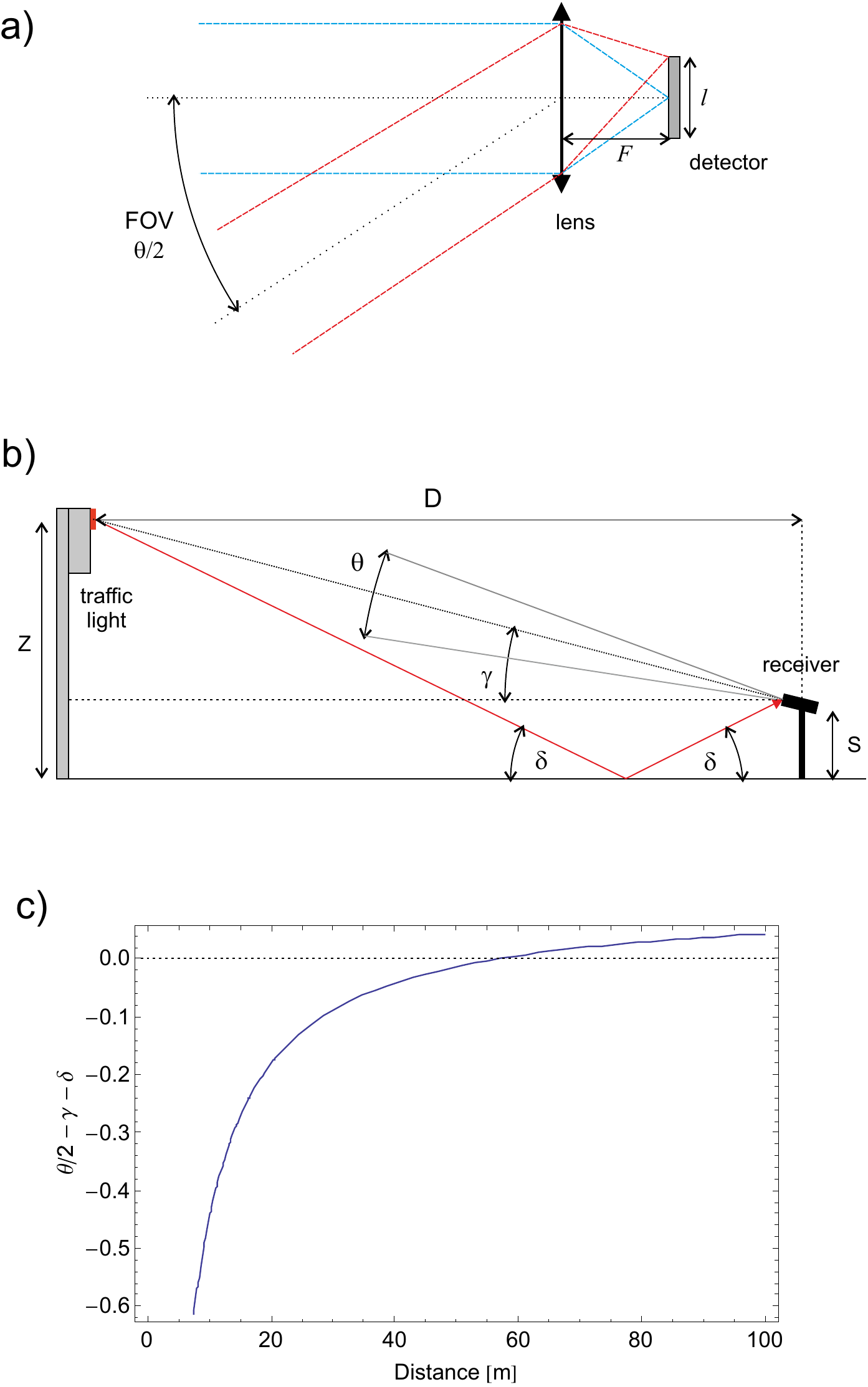}
    \caption{a) Field of View for a given detector of size $l$ and focal length $F$; b) reflections form tarmac (red line) in relation to parameters used in the text; c) the merit function $\theta/2 - \gamma - \delta$ gets positive when reflections can geometrically enter the detector's FOV.}
    \label{fig:FOV}
\end{figure}

Here we show that in our configuration reflections of lamp's light from tarmac are never collected by receiver as they fall out of  the Field of View  (FOV) of the receiver anywere on the measurements grid. Without loss of generality we restrict our analysis to the case of receiver placed along the optical axis of the traffic light lens ($\beta=0$). With reference to Fig.\,\ref{fig:FOV}-a, the FOV $\theta$ of a receiver is given by $\theta$/2 = arctan$(l/2F)$, where $l$ and $F$ are the detector width and lens focal length, respectively.
As shown in Fig.\,\ref{fig:FOV}-b, one can see that reflections from tarmac (red line) cannot geometrically enter the detector FOV when $\theta/2\leq \gamma + \delta$.
After easy trigonometric considerations, the above expression reads as:
\begin{equation}
    \mathrm{arctan}\left(\frac{l}{2F}\right)\leq\mathrm{arctan}\left(\frac{Z+S}{D}\right) + \mathrm{arctan}\left(\frac{Z-S}{D}\right) ,
\end{equation}
where the parameters are described in Fig.\,\ref{fig:FOV}).
The above expression, given our parameters set ($F=30$\,mm, $l=6$\,mm) is always verified in our measurement grid ($D<35$\,m) and for all of the three receiver heights, and turns false only at very large distances ($D\gtrsim50$\,m, see Fig.\,\ref{fig:FOV}-c), where the detector's optical axis progressively leans towards a horizontal plane as the distance $D$ grows, hence letting to reflected beams the possibility to enter the FOV cone.

\end{document}